%% file: 00-all.tex
\documentclass[11pt]{article}
\usepackage{amsmath, amssymb, amsthm, soul}
\usepackage[numbers, square]{natbib}
\usepackage[left=1in, bottom=1in, right=1in, top=1in]{geometry}
\usepackage[citecolor=blue, urlcolor=blue]{hyperref}
\usepackage{dsfont}
\usepackage{xcolor}
\usepackage{natbib}
\usepackage{multicol}
\usepackage{multirow}
\usepackage{graphicx}
\usepackage{subcaption}
\usepackage{caption}
\usepackage{comment}
\usepackage{wrapfig}
\usepackage{mathtools}
\usepackage{booktabs}
\usepackage{bbm}
\DeclarePairedDelimiterX{\norm}[1]{\lVert}{\rVert}{#1}
\usepackage{subcaption}

\definecolor{applegreen}{rgb}{0.55, 0.71, 0.0}
\DeclareMathOperator*{\argmin}{argmin}

\usepackage{wrapfig} 
\usepackage[font=small,skip=0pt]{caption}
\newtheorem{theorem}{Theorem}
\usepackage{authblk}

\title{BLOG: Bayesian Longitudinal Omics with Group Constraints}

\author[1]{Livia Popa\thanks{Corresponding author: \texttt{lp472@cornell.edu}}}
\author[1]{Sumanta Basu}
\author[2]{Myung Hee Lee}
\author[1]{Martin T. Wells}

\affil[1]{Department of Statistics and Data Science, Cornell University, New York, USA}
\affil[2]{Center for Global Health, Department of Medicine, Weill Cornell Medicine, New York, NY}

\date{ }

\begin{document}

\maketitle

\begin{abstract}
%include  1 sentence answers to each of the bullet point questions.
Clinical investigators are increasingly interested in discovering computational biomarkers from short-term longitudinal omics data sets. This work focuses on Bayesian regression and variable selection for longitudinal omics datasets, which can quantify uncertainty and control false discovery. In our univariate approach, Zellner's $g$ prior is used with two different options of the tuning parameter $g$: $g=\sqrt{n}$ and a $g$ that minimizes Stein's unbiased risk estimate (SURE). Bayes Factors were used to quantify uncertainty and control for false discovery. In the multivariate approach, we use Bayesian Group LASSO with a spike and slab prior for group variable selection. In both approaches, we use the first difference ($\Delta$) scale of longitudinal predictor and the response. These methods work together to enhance our understanding of biomarker identification, improving inference and prediction. We compare our method against commonly used linear mixed effect models on simulated data and real data from a Tuberculosis (TB) study on metabolite biomarker selection. With an automated selection of hyperparameters, the Zellner's $g$ prior approach correctly identifies target metabolites with high specificity and sensitivity across various simulation and real data scenarios. The Multivariate Bayesian Group Lasso spike and slab approach also correctly selects target metabolites across various simulation scenarios.
\end{abstract}

\input{01-intro}

\input{02-background}

\input{03-method}

\input{04-simulation}
\input{05-realdata}
\input{06-conclusion}

\newpage 

\bibliographystyle{abbrvnat}
\bibliography{biblio-bayesprolong}

\newpage
\appendix
\input{07-appendix}

\end{document}

%% file: 01-intro.tex
\newpage

\section{Introduction}\label{sec:intro}

% problem statement
Clinical investigators have become increasingly interested in collecting data on a large number of omics features in order to build computational biomarkers that can predict the clinical outcome. These predictions can serve as indirect indicators of disease progression and can provide new information about physiological processes \cite{dutta_integration_2020, cliff_distinct_2013,li_longitudinal_2021}. 

Existing methods in the high-dimensional setting use penalized regression and do not offer uncertainty quantification. This work focuses on Bayesian high-dimensional regression and variable selection for longitudinal omics datasets, which can quantify uncertainty, control for false discovery, and addresses the limited power in short-term longitudinal studies. To address the challenges of high-dimensional longitudinal omics data, we present two complementary Bayesian approaches—one univariate and one multivariate. The univariate method offers interpretable, marginal evidence for individual biomarker associations, while the multivariate method captures joint effects and correlation structures across features. Presenting both approaches in a single framework provides versatile tools for clinical investigators that are tailored to different inference goals and study designs such as identifying individual associations or capturing joint effects across features.

By operating on the first difference scale, the projected model effectively controls for time-invariant heterogeneity and captures dynamic biomarker-outcome associations over time. These extensions use uncertainty quantification and mixed-effects modeling using a combination of Bayesian regression and regularization. This work makes significant methodological contributions to the analysis of high-dimensional longitudinal omics data, with direct relevance to biomarker discovery. 

% significance
This work is motivated by two clinical trial studies. The first is a study of tuberculosis consisting of 4
time point observations for 15 subjects each with a measured mycobacterial load of the sputum, quantified by the time in hours to the positivity of Mtb culture (TTP),
and abundances of up to 915 urinary metabolites. We also evaluated our method in an intensive lifestyle intervention study aimed at understanding the underlying reversible causes of diabetes among patients \cite{taheri2020diadem}. This study contained 5 time point observations for 16 subjects, each with measured change in three outcomes (HDL, I2GDF15, and I2MIC1) and 17 protein abundances.

Previous work has been
performed on the identification of metabolite-based biomarkers for
patients with tuberculosis \cite{PROLONG, ISA2018157,10.1172/jci.insight.136301} and the identification of biomarkers affecting diabetic patients \cite{ramel2013neonatal, ha2023one}. 
% rigor of prior research
Existing work in the clinical omics literature uses univariate longitudinal predictive regression models \cite{walsh_early_2020, wipperman_gastrointestinal_2021} for the outcome of interest. In order to select significant variables, these methods compare p-values across univariate models. These models lack consideration for correlations of features over time which leads to reduced sensitivity and reduced statistical power.

The literature on high-dimensional omics in molecular biology and genomics is vast, but these methods are not designed around using a time-varying clinical outcome to guide variable selection. The most common method for analyzing such data sets is linear mixed-effects model to regress clinical outcomes on one or more feature trajectories, but these methods are often not powerful enough unless data is collected
over a moderate time horizon. More recently, the PROLONG (Penalized regression on longitudinal omics data with network and group constraints) method was proposed \cite{PROLONG}, which can jointly select longitudinal features that co-vary with a time-varying outcome on the first difference scale. This method leverages correlation between variable trajectories by using the first-difference scale, which controls for time dependence within the data. This improves power and eliminates the need for time-related assumptions. PROLONG also introduces network and group lasso constraints to induce sparsity, while also grouping different time components of each metabolite and incorporating the correlation structure of the data. The current version of PROLONG does not offer uncertainty quantification, which is important for precisely controlling false discoveries and assessing the statistical significance of biomarkers.

% description of approach
This work focuses on uncertainty quantification and controlling for false discovery using Bayesian high-dimensional regression and variable selection for longitudinal omics datasets. In the Bayesian setting, we adopt an empirical Bayesian approach as well as hierarchical Bayes principles for hyperparameter selection. In the univariate setting, Zellner-g prior is used to generate posterior distributions to obtain Bayes factors to quantify uncertainty. Two choices of $g$ are considered to highlight the trade-off between empirical optimality and practical performance. The first choice being one that minimizes Stein's unbiased risk estimate (SURE), offering data-driven calibration, while the second, $g= \sqrt{n}$,has shown strong empirical performance and competitive variable selection behavior comparable to LASSO. Including both allows us to assess robustness across different prior specifications and offers practical guidance for clinical investigators. Bayes factors were then calculated for both choices of $g$ using a Bayes factor proposed in \cite{liang2008mixtures}. More details are given in the Methods Section \ref{sec: method}. 

%In a simulation study, when compared to other choices of $g$ in a linear regression setting, $g= \sqrt{n}$ outperforms others in terms of RMSE for point estimation, MIS for interval estimation, MIS for interval prediction, RMSE for point prediction, 1-AUPRC, and CPU time (in comparison to LASSO) \cite{doi:10.1073/pnas.2120737119}.

%In the null-based approach to calculating Bayes factors, each model $\mathcal{M}_\gamma$ is compared with the null model $\mathcal{M}_N$ through the hypotheses $H_0:\beta_\gamma=0$ and $H_1:\beta_\gamma \in \mathbb{R}^{p_\gamma}$  where the resulting Bayes factor for comparing any model $\mathcal{M}_\gamma$ to the null model $\mathcal{M}_N$ is: \begin{equation}
%    \text{BF}[\mathcal{M}_\gamma:\mathcal{M}_N]=(1+g)^{(n-p_\gamma-1)/2}[1+g(1-R^2_{\gamma})]^{-(n-1)/2}
%\end{equation}
%Where $R^2_\gamma$ is the ordinary coefficient of determination of the OLS regression model, $g$ is the particular choice of $g$ in calculating Zellner's g prior, and $p_\gamma$ is the number of predictors in the model. {\color{red}{[SB: consider reducing technical details in the introduction.]}}

In the multivariate case, we use a Bayesian formulation of Group LASSO with spike and slab priors \cite{xu2015bayesian}. This procedure uses a multivariate point mass mixture similar to Zhang et al. (2014)\cite{10.1111/rssc.12053} that facilitates the selection of group variables by producing exact 0 estimates at the group level. We evaluated this method using posterior mean thresholding, which can select and estimate variables at the same time. 

% results of validation (one paragraph each on theory, sim and realdata)
To validate these methods, we simulate target and noise variables using different signal-to-noise ratios and compare them to the following benchmark methods presented in the PROLONG paper: Linear Mixed Effect model (LME) p-values and the Univariate PROLONG Wald test p-values \ref{sec:sim_univ}. We find that the univariate model correctly identifies target variables from noise and quantifies uncertainty by giving significant Bayes factors based on thresholding described in Section \ref{sec: method}. We also find that the univariate method is able to identify significant metabolites from the TB data and give corresponding significant Bayes factors compared to the PROLONG Wald tests. The univariate method can also identify proteins of interest to clinical investigators when applied to diabetes data \ref{sec:realdata}.

For the multivariate method, we simulate the same signal-to-noise ratios and find that the method is able to correctly distinguish targets from noise. The model also provides nonzero posterior medians for targets and zero medians for noise variables which are used to quantify uncertainty. The use of posterior median thresholding is discussed in the Methods Section \ref{sec: method}.

A central innovation is the application of Zellner's $g$-prior and the Group LASSO with spike and slab prior within this framework to enable rigorous variable selection through the computation of Bayes factors. This allows for formal uncertainty quantification and improved false discovery control in high-dimensional settings, addressing a long-standing challenge in genomic data analysis. It is demonstrated that the Bayes approaches have strong performance in terms of sensitivity and specificity across simulation scenarios.

% summary of contributions: novelty
In this research, we introduce a novel approach by being the first to conduct Bayes factor-type evaluations within a Bayesian framework as part of an effort to quantify evidence for metabolite-drug and protein-outcome interactions over time. This method extends the capabilities of the PROLONG method, reinterpreting through a Bayesian lens, which allows for more flexible and robust uncertainty quantification. Furthermore, we applied a Bayesian group lasso method to these unique datasets, leveraging its sparsity-inducing properties to identify key metabolites with varying responses to drug intervention. A critical contrast emerges between the BLOG and group lasso approaches: While BLOG aims to provide probabilistic assessments of model evidence, the group lasso focuses on selecting significant variables and estimating their effects with precision. Together, these methodologies complement each other, allowing us to deepen our understanding of the behavior of metabolites in response to the drug, improving both inference and predictive accuracy.

%{\color{red}{[SB: novelty is inference + FDR all at once using appropriate Bayesian methods.]}}

% outline of the rest of the paper
The paper is organized as follows. In the background Section \ref{sec:background}, we review the relevant literature on high-dimensional regression for longitudinal omics data and discuss the limitations of current models, such as PROLONG, which lack uncertainty quantification. In the method Section \ref{sec: method}, we introduce our Bayesian approach, which uses Zellner's $g$-prior in the univariate case and a Bayesian group LASSO method in the multivariate case to enhance biomarker discovery by quantifying uncertainty and controlling false discoveries. The simulation Section \ref{sec:simulation} evaluates the performance of our method across various signal-to-noise ratios, contrasting its effectiveness in identifying target variables compared to traditional models, and assessing the impact of different tuning parameter choices on model performance. In the real data Section \ref{sec:realdata}, we apply our method to the two mentioned datasets. These applications demonstrate our method's ability to identify significant biomarkers in complex, longitudinal datasets. Finally, the conclusion \ref{sec:conclusion} summarizes our findings, highlighting the advantages of using Bayesian methods for uncertainty quantification in high-dimensional settings and discussing potential directions for future research.

%% file: 02-background.tex
\section{Background}\label{sec:background}
To provide motivation for this Bayesian high-dimensional regression and variable selection, we first describe the mixed univariate models often used for longitudinal omics regression \cite{walsh_early_2020} as well as the PROLONG \cite{PROLONG} method, which provides an alternative model and a natural hypothesis testing technique, as well as a multivariate extension that includes constraints to induce sparsity in the high-dimension scenario to incorporate the dependence structure of omics predictors. We will also provide a background of previous Bayesian approaches to this problem in the univariate and multivariate case.

\subsection{Linear Mixed Effect Models}
Linear Mixed Effect Models are used frequently when identifying biomarkers of TB (see \cite{adekambi2015biomarkers, bauer2015health}) and diabetes (see \cite{ha2023one, ramel2013neonatal}) biomarkers. In our case, the features described are biomarkers. We analyze data from a short-term tuberculosis drug trial and a short-term diabetes lifestyle intervention study. The data from the short-term trials consist of regressors $\tilde{X}_{it}^{[j]}$ and responses $\tilde{Y}_{it}$ for subject $i=1,...,n$, feature $j=1,...,p$, and time point $t=1,...,T$. A univariate linear mixed effects model for the levels of $\tilde{Y}_{it}$ with fixed effects for the feature and for the time component and a subject-specific fixed or random effect takes the form, \begin{equation}
\label{model1}
\tilde{Y}_{it}=\alpha_t^{[j]}+\beta_t^{[j]}\tilde{X}_{it}^{[j]}+b_i+\epsilon_{it},
\end{equation}
where $\alpha_t^{[j]}$ is the mean response across subjects at time $t,b_i$ is the subject specific fixed or random effect, $\epsilon_{it}$ is the mean-zero error term and $\beta_t^{[j]}$ is the time-dependent feature-specific parameter of interest \cite{wooldridge2010econometric}. The choice of fixed- and random-effects models depends on the unobserved differences in the panel data. If unobserved factors that are controlled for unchanging differences are suspected to be correlated with the explanatory variables, then a fixed effects model may be preferred. Otherwise, if the variation across individuals is more of an interest, a random-effects model may be more suitable. If the effects for each individual are significantly correlated with the explanatory variables, then using a random effects model can lead to biased and inconsistent estimates \cite{cornell2014random}. 

Longitudinal fixed- and random-effects models are used to address changes in subjects over time, particularly focusing on short-term changes and trends in feature levels. In scenarios where time-dependent factors are crucial \cite{wooldridge2010econometric}, such models lack enough power to accurately capture true features with short-term changes. LME models are inherently doing separate univariate modeling of all features, and a joint model such as PROLONG \cite{PROLONG}, holds the promise to increase power. In addition to the increased power, our BLOG method is able to quantify uncertainty and provide a more powerful inference procedure. 
\subsection{PROLONG} 
% describe necessary maths/algorithmic background to completely state the problem rigorously

To improve power, a first difference approach is proposed in \cite{PROLONG}, where changes between consecutive time points are emphasized. Longitudinal data time points are taken in somewhat close increments, where the data is positively correlated. By regressing the first differences of the outcomes on the first differences of features, PROLONG aims to identify features that change over time between subjects in a way consistent with the changes in the outcomes. This approach works better in the same way that a paired t-test is more powerful than an unpaired t-test when there is time-invariant heterogeneity. Furthermore, the first-difference approach helps mitigate potential endogeneity concerns related to time-invariant unobserved factors \cite{PROLONG}. In the case of both of our real data applications, when subjects display different responses to treatment over time, identifying significant linear relationships on this delta scale can provide insight into biomarkers that could affect treatment outcomes.

\subsubsection{Univariate PROLONG}

In the univariate case, we set up the matrix representation of the differenced regression models in the same way as the univariate case of PROLONG \cite{PROLONG}. For each subject, all time points are matricized to be incorporated into a single model. Once the time points are in matrix form, we represented them in terms of other matrices in order for them to be referenced on the delta scale.

Given the original model in \eqref{model1}, the $n \times T$ response matrix $\left( \left(\tilde{Y}_{it} \right) \right)_{1 \le i \le n, 1 \le t \le T}$, 
\iffalse
can be represented and differenced in the following step 
\begin {equation*}
\begin{bmatrix}
    \tilde{Y}_{11} & \cdots &  \tilde{Y}_{1T} \\
    \multicolumn{3}{c}{$\vdots$}    \\  
\tilde{Y}_{n1} & \cdots & \tilde{Y}_{nT}
\end{bmatrix}_{n \times T} \;
\rightarrow \; \;
\begin{bmatrix}
    \Delta \tilde{Y}_{11} & \cdots & \Delta \tilde{Y}_{1(T-1)} \\
    \multicolumn{3}{c}{$\vdots$}    \\  
 \Delta \tilde{Y}_{n1} & \cdots & \Delta \tilde{Y}_{n(T-1)}
\end{bmatrix}_{n \times (T-1)}
\end{equation*}
\fi
denote $\Delta \tilde Y_{it}$ as the first-difference  $\tilde Y_{i(t+1)} - \tilde Y_{it}$, for $i = 1, \ldots, n$ and $t = 1, \ldots, T-1$.
Vectorizing this $n \times (T-1)$ matrix of $\Delta \tilde Y_{it}$ values yields the ${n(T-1) \times 1}$ vector $Y$ to be used as the dependent variable in the differenced model
\begin {equation}
\label{depvar}
Y = [ \Delta \tilde{Y}_{11}, \hdots, \Delta \tilde{Y}_{n1} \hdots, \Delta \tilde{Y}_{1(T-1)},  \hdots, \Delta \tilde{Y}_{n(T-1)} ]^\top.
\end{equation}

Similarly, for the $j^{\text{th}}$ feature , $j=1, \ldots, p$, the $n \times T$ matrix of predictors $\left( \left(\tilde{X}^{[j]}_{it} \right) \right)_{1 \le i \le n, 1 \le t \le T}$
can be differenced 
\iffalse
in the following step
\begin{equation*}
\begin{bmatrix}
\tilde{X}^{[j]}_{11} & \cdots & \tilde{X}^{[j]}_{1T} \\
    \multicolumn{3}{c}{$\vdots$}    \\  
\tilde{X}^{[j]}_{n1} & \cdots & \tilde{X}^{[j]}_{nT} 
\end{bmatrix}_{n \times T}
\rightarrow \; \;
\begin{bmatrix}
\Delta \tilde{X}^{[j]}_{11} & \cdots &  \Delta \tilde{X}^{[j]}_{1(T-1)} \\
    \multicolumn{3}{c}{$\vdots$}    \\  
\Delta \tilde{X}^{[j]}_{n1} & \cdots & \Delta \tilde{X}^{[j]}_{n(T-1)} 
\end{bmatrix}_{n \times (T-1)}
\end{equation*}
\fi
to form a $n \times (T-1)$ matrix with entries  $\Delta \tilde X_{it}^{[j]}$, the first-difference  $\tilde X_{i(t+1)}^{[j]} - \tilde X_{it}^{[j]}$.

To derive the difference design matrix we need to respect the temporality of the observations and regress $\Delta \tilde{Y}_{ik}$ for $k = 1, \ldots, T-1$ on the differenced measurements of $\Delta \tilde{X}^{[j]}_{it}$ for $t=1, \ldots, k$  and $j=1, \ldots, p$. % we do not want to regress $\Delta \tilde{Y}_{t}$ on later measurements of $\Delta \tilde{X}$.  
 Consequently, for the $j^{\text{th}}$ feature, the $n(T-1) \times T(T-1)/2$ design matrix is

% \setlength{\arraycolsep}{1pt}
%\resizebox{.945\linewidth}{!}{%
%$\displaystyle
\begin{equation}
\label{indvar}
 X^{[j]} =
\left[
\begin{array}{c|c|c|c}  
  \Delta \tilde{X}^{[j]}_{11} & \multirow{3}*{0} &   \multirow{3}*{0} &  \multirow{3}*{0} \\
\vdots &   & &    \\
  \Delta \tilde{X}^{[j]}_{n1}   &   & & \\
\hline
\multirow{3}*{0} & \Delta \tilde{X}^{[j]}_{11} \quad \Delta \tilde{X}^{[j]}_{12} &   \multirow{3}*{0} &   \multirow{3}*{0} \\
 & \vdots  & &  \\
&  \Delta \tilde{X}^{[j]}_{n1} \quad \Delta \tilde{X}^{[j]}_{n2} & &  \\
\hline 
0 &  0 &  \ddots & 0 \\
\hline
  \multirow{3}*{0} &   \multirow{3}*{0} &  \multirow{3}*{0} & \Delta \tilde{X}^{[j]}_{11} \quad \cdots \quad \Delta \tilde{X}^{[j]}_{1(T-1)}  \\
 &   & & \vdots \\
&   & &   \Delta \tilde{X}^{[j]}_{n1} \quad \cdots \quad \Delta \tilde{X}^{[j]}_{n(T-1)} 
\end{array}
\right].
\end{equation}
%$}.
Now the model is \begin{equation*}
    Y =  X^{[j]}\beta^{[j]} + \epsilon ,
\end{equation*}
where $Y$ and $X^{[j]}$ are described above \eqref{depvar}, \eqref{indvar}, $\epsilon$ is some mean zero error term and $\beta^{[j]} $ corresponds to the $jth$ feature which we are testing whether or not is a significant biomarker. This model is equivalent to a the levels model in (\ref{model1}) with a constant intercept. A proof of this is given in the appendix \ref{sec:appendix}. 

Based on this set-up, PROLONG and Wald test is able to give a rank list of significant metabolites and proteins but now with our univariate BLOG we can also produce Bayes factors to quantify uncertainty and control for false discovery. PROLONG uses false discovery rate (FDR) correction, but they do not assume a joint model. With our Bayesian method, by incorporating priors we can generate posterior distributions to have more powerful biomarker discovery. The application of Zellner's $g$-prior to this matrix setup is described in the methods Section \ref{sec: method}.
% describe significance of the problem in detail - cite papers where these methods were used or were proven to be useful

% describe rigor of prior research - what remains lacking? Did anyone else talk about it? This is also the place where you describe your potential benchmark methods.

% in your opinion, why is filling this gap non-trivial. e.g. why some immediate/obvious approaches won't work? This can be written in the end

\subsubsection{Multivariate PROLONG}
The PROLONG model uses the univariate regression structure, but incorporates all $\beta^{[j]}$'s simultaneously for a joint model. The exact structure is described in greater detail in the PROLONG article \cite{PROLONG} but the general model is \begin{equation*}
    Y = X\beta + \epsilon,
\end{equation*} where X has the same matrix structure as \eqref{indvar} but now the $\Delta \tilde{X}^{[j]}_{ik}$ in \eqref{indvar} is replaced with row vector \begin{equation*}
    \Delta \tilde{X}^{[j]}_{ik} = \left[ \Delta \tilde{X}^{[1]}_{ik}, \Delta \tilde{X}^{[2]}_{ik},\cdots, \Delta \tilde{X}^{[p]}_{ik}\right],
\end{equation*} leading to the differenced $n(T-1) \times pT(T-1)/2$ design matrix. Our matrix structure is described in the methods Section \ref{sec: method}

One of the tuning parameters of multivariate PROLONG is a group lasso penalty. This is a generalized version of LASSO (Yuan and Lin, 2006 \cite{yuan2006model}) that is able to select grouped variables in order to make accurate regression predictions. From our mixed effect regression problem stated above \eqref{model1}, we can obtain the group lasso estimator by solving
\begin{equation*}
    \min_{\beta} \bigg\| Y - \sum_{j=1}^{p} X^{[j]} \beta^{[j]} \bigg\|_2^2 + \lambda \sum_{j=1}^{p} \| \beta^{[j]} \|_2,
\end{equation*}
where $P$ is the number of groups of variables.

Along with frequentist group lasso penalty, PROLONG uses a network constraint via the Laplacian matrix of the graph associated with a dependence matrix, but we do not pursue this direction in our work.

%% file: 03-method.tex
\section{Method: BLOG}\label{sec: method}

We extend the original PROLONG model for high-dimensional longitudinal data by introducing a Bayesian framework that leverages appropriately specified priors for first-difference models. This enhancement facilitates rigorous inference and uncertainty quantification, with particular utility in biomarker discovery. Specifically, we examine both univariate and multivariate approaches: the univariate formulation employs the Zellner-$g$ prior, for which we detail the choices of $g$ and the Bayes factor thresholds used in model validation; the multivariate formulation utilizes the Bayesian group lasso, where posterior medians are employed for uncertainty quantification and validation purposes.

% describe your proposed method/approach rigorously. It should have all the details that a reader will need to reproduce/implement your method. It should also clearly motivate the purpose of each step/part of the method.

\subsection{Univariate BLOG}

In this Bayesian regression setup, we use the normal-inverse gamma model for $\beta$ and $\sigma^2$. This model allows for flexible modeling of the prior mean and covariance matrix of $\beta$. The normal-inverse gamma model specifies $\beta | \sigma^2 \sim N(\beta_0, \sigma^2 V_0)$ and $\sigma^2 \sim \Gamma^{-1}(a,b)$ where $\beta_0 \in \mathbb{R}^{p x 1}, V_0 \in \mathbb{R}^{p x p}, a,b > 0$. From this setup, it is said that $\beta, \sigma^2$ both follow a normal-inverse gamma with parameters $(\beta_0, V_0, a,b)$ which is a conjugate prior. Then the posterior distribution of $(\beta , \sigma^2)$ is also normal-inverse gamma with parameters $(\beta_*, V_*,a_*,b_*)$ which are defined as:
\begin{equation*}
    \beta_* = (V_0^{-1}+X^TX)^{-1}(V_0\beta_0 + X^TY)
\end{equation*}
\begin{equation*}
    V_* = (V_0^{-1}+X^TX)^{-1}
\end{equation*}
\begin{equation*}
    a_* = a + \frac{n}{2}
\end{equation*}
\begin{equation*}
    b_* = b + \frac{1}{2}(\beta^T_0 V^{-1}_0 \beta_0 + Y^TY-\beta^T_*V^{-1}_*\beta_*).
\end{equation*}
The conditional posterior of $\beta|\sigma^2$ and the posterior of $\sigma^2$ are $\beta | \sigma^2, Y\sim N(\beta_*,\sigma^2V_*)$ and $\sigma^2 | Y \sim \Gamma^{-1}(a_*,b_*)$. The posterior mean $\beta_*$ are the estimated regression parameters which in our case are the $p_*=6$ estimated coefficients for each biomarker. If at least one of these coefficients is nonzero, we consider the particular biomarker to be significant.

When defining $\beta_0$ in our model, we wish to place priors with zero mean on the $p=1,..,6$ time points suggesting that there is no signal. Noting for the linear model $Y = X\beta + \epsilon$ where $\epsilon \sim N(0,\sigma^2 I_n)$, the least-squares estimator $\hat{\beta}_{OLS}=(X^TX)^{-1}X^TY$ has the covariance matrix $\sigma^2(X^TX)^{-1}.$ We choose $V_0 = g(X^TX)^{-1}$, where $g>0$. This choice is a special case of the normal-inverse gamma model known as \textit{Zellner's g-prior} \cite{Zellner1986}. Using this choice of $V_0$ when calculating the posterior mean $\beta_*$ in \ref{eq:betastar} and the covariance matrix $V_*$ in \ref{eq:vstar}, we obtain 
\begin{equation}
    \beta_* = \frac{1}{1+g} \beta_0 + \frac{g}{1+g} \hat{\beta}_{OLS}
    \label{eq:betastar}
\end{equation}
\begin{equation}
    V_* = \frac{g}{1+g}(X^TX)^{-1}
    \label{eq:vstar}.
\end{equation}

The scalar hyperparameter \( g \) controls the strength of the prior: small values imply strong regularization, while large values render the prior diffuse. However, when used with a fixed and large \( g \), Zellner's prior performs poorly from an information-theoretic perspective. A major concern is \textit{Bartlett's paradox}: as \( g \to \infty \), the marginal likelihood for complex models shrinks toward zero, regardless of data quality. Consequently, the Bayes factor always favors the null model, even when the data provide strong evidence for the alternative \citep{bartlett1957paradox}. A related phenomenon is \textit{Lindley's paradox}, where posterior model probabilities overwhelmingly favor the null model as the sample size increases, despite strong likelihood support for more complex models \citep{lindley1957paradox}.

These paradoxes are especially problematic in high-dimensional settings, where the number of parameters \( p \) grows with or exceeds the number of observations \( n \). In such regimes, the marginal likelihood under a fixed-\( g \) prior decreases exponentially in \( p \), because the volume of parameter space is poorly supported by the prior. This leads to a systematic underestimation of the complexity of the model, even when relevant covariates are present in the data. Furthermore, fixed \( g \) prior lacks the adaptivity needed to balance signal detection and overfitting in sparse or poorly posed problems, common in modern machine learning and genomics.

Choosing the tuning parameter $g$ is important task in our method. We experiment with two potential choices of $g$ that exist in the literature. \smallskip

\noindent \textbf{\textit{Choice of $g$ with SURE Minimization}}. 
%In the univariate case, there were two choices of g that were chosen when using Zellner's g-prior. 
The first choice of $g$ was motivated by \cite{zellner}. In their article, they found that the optimal choice of g was to minimize Stein's unbiased risk estimate (SURE), an unbiased estimate of $||\hat{Y}-X\beta||^2$. They first suggested a method to select the value of $g$ that minimizes the sum of squared residuals $||Y- \hat{Y}||^2$, where $\hat{Y}=X\beta_*$ is the vector of fitted values: \begin{equation*}
    \begin{aligned}
    \label{resid}
    \hat{Y} = X\beta_* &= X(\frac{1}{1+g}\beta_0 + \frac{g}{1+g} \hat{\beta}_{\text{OLS}})\\
    &= \frac{1}{1+g}Y_0 + \frac{g}{1+g} \hat{Y}_{\text{OLS}}
    \end{aligned}
\end{equation*} 
where $Y_0=X\beta_0$ and $\hat{Y}_{\text{OLS}}=X\hat{\beta}_{\text{OLS}}$. They then found that there were no analytical solutions to $g_* = \argmin_{g>0}||Y-\hat{Y}||^2$. Instead, they chose to minimize Stein's unbiased risk estimate (SURE) \cite{fourdrinier2012improved, fourdrinier2018shrinkage}, which is an unbiased estimate of $||\hat{Y}-X\beta||^2$. More details are found in Theorem \ref{thm:surethm1}, stated in the Appendix. They also showed that the value of $g$ that minimizes the SURE estimate is $g = \frac{||\hat{Y}_{\text{OLS}}-Y_0||^2}{p_* \hat{\sigma}^2}-1$, as stated in the Appendix  \ref{thm:surethm2}.

\iffalse
\begin{theorem} \label{thm:surethm1}
(SURE for linear models). Let $Y\sim N(X\beta,\sigma^2 I_n)$, where the dimensions of $X$ are $n \times p$, and let $\hat{\beta} = \hat{\beta}(Y)$ be a weakly differentiable function of the least squares estimator $\hat{\beta}_{\text{OLS}}$ such that $\hat{Y} = X\hat{\beta}$ can be written in the form $\hat{Y}=a+SY$ for some vector \textbf{a} and matrix \textbf{S}. Let $\hat{\sigma}^2=||Y-X\hat{\beta}_{\text{OLS}}||^2/(n-p)$. Then,\begin{equation}
    \delta_0(Y) = ||Y-X\hat{\beta}||^2+(2\text{Tr}(S)-n)\hat{\sigma}^2
\end{equation}
is an unbiased estimator of $||Y-X\hat{\beta}||^2$.
\end{theorem}
\begin{theorem} \label{thm:surethm2}
(SURE minimization with respect to $g$). The value of $g$ that minimizes SURE in \eqref{resid} is \begin{equation}
    \label{gsure}
    g_* = \frac{||\hat{Y}_{\text{OLS}}-Y_0||^2}{T \hat{\sigma}^2}-1
\end{equation}
\end{theorem}
\fi
This value of $g$ was used in the univariate case of this model, where $p_*$ is the number of differenced time points to be estimated for each biomarker. In our matrix set-up above, we have $p_*=T(T-1)/2$ time points with each one being an entry in the $\beta_*$ vector of dimension $p_* \times 1$. \smallskip

\noindent \textbf{\textit{Choice of} $g = \sqrt{n}$}. 
Attempts to mitigate the problems by scaling \( g \) with the sample size may reduce some of the paradoxes and artifacts \citep{bayarri2012criteria}.  Consequently. another choice of $g$ that was used in the univariate case was $g = \sqrt{n}$, where $n$ is the number of observations. In the TB case, subjects $n = 15$. This choice of $g$ is motivated by \cite{doi:10.1073/pnas.2120737119}. In this article, it was found that when using a probability model for parameter estimation, interval estimation, inference about model parameters, point prediction, and interval prediction, focusing on variables to include in a linear regression model, three adaptive Bayesian model averaging methods outperformed other models across all mentioned statistical tasks. 

%In a simulation study, when compared to other choices of $g$ in a linear regression setting, $g= \sqrt{n}$ outperforms others in terms of RMSE for point estimation, MIS for interval estimation, MIS for interval prediction, RMSE for point prediction, 1-AUPRC, and CPU time (in comparison to LASSO) \cite{doi:10.1073/pnas.2120737119}. 

These models used Zellner's $g$-prior for the parameters and among the three, $g=\sqrt{n}$ and an empirical Bayes-local prior were shown to be competitive with the least absolute shrinkage and selection operator (LASSO). Since we focus on a choice of $g$ for Zellner's $g$ prior, $g=\sqrt{n}$ was chosen rather than the empirical Bayes local prior.

\subsubsection{Bayes Factor Thresholds}
For both choices of $g$, Bayes factors were calculated using a method proposed in \cite{liang2008mixtures}. In the null-based approach to calculating Bayes factors, each model $\mathcal{M}_\gamma$ is compared with the null model $\mathcal{M}_N$ through the hypotheses $H_0:\beta_\gamma=0$ and $H_1:\beta_\gamma \in \mathbb{R}^{p_\gamma}$  where the resulting Bayes factor for comparing any model $\mathcal{M}_\gamma$ to the null model $\mathcal{M}_N$ is: \begin{equation}
    \text{BF}[\mathcal{M}_\gamma:\mathcal{M}_N]=(1+g)^{(n-p_\gamma-1)/2}[1+g(1-R^2_{\gamma})]^{-(n-1)/2}.
\end{equation}
Where $R^2_\gamma$ is the ordinary coefficient of determination of the OLS regression model, $g$ is the particular choice of $g$ in calculating Zellner's g prior, and $p_\gamma$ is the number of predictors in the model. We quantify uncertainty by providing Bayes factors for each variable. These factors summarize the evidence provided by the data in favor of one scientific theory, represented by a statistical model (Kass and Raftery 1995) \cite{kass1995bayes}. We consider twice the natural logarithm of the Bayes factor, which is on the same scale as the familiar deviance and likelihood ratio test statistics. Bayes factor thresholds and interpretations are as follows: 
\begin{table}[h]
\centering
\begin{tabular}{ccc}
$2 \log_e(B_{10})$ &\textbf{Bayes Factor }$(B_{10})$ & \textbf{Evidence against $H_0$ }\\
0 to 2 &1 to 3 & Not worth more than a bare mention\\
2 to 6&3 to 20 & Positive\\
6 to 10&20 to 150 & Strong \\
$> 10 $ & $> 150$ & Very Strong\\
\end{tabular}
\caption{Table for Bayes Factor and Evidence against the null hypothesis}
\label{tbl:BF}
\end{table}

A very strong Bayes factor in the Bayesian setting corresponds to a significant p-value and FDR rate in the frequentist setting. Empirical false-positive and true-positive rates are reported, as well as FD rates.

\subsection{Multivariate BLOG} \label{sec:MVM}

Although Zellner's $g$ prior worked well in the univariate setting, existing high-dimensional Zellner's priors \ref{app:zellgpgreatern} did not work in our set up, and we adopted a Bayesian group lasso approach for the multivariate problem.

Using the proposed regression structure from the univariate case, we use a matrix setup similar to multivariate PROLONG. We now create a matrix X that contains all $X^{[j]}, (j=1,...p)$ consecutive metabolites. The $n(T-1) \times pT(T-1)/2$ design matrix is now:
\begin{equation}
\label{multixmatrix}
 X =
\left[
\begin{array}{c|c|c|c}  
X^{[1]} & X^{[2]}  & \hdots & X^{[p]} \\
\end{array}
\right].
\end{equation}

BLOG uses a Bayesian formulation of group lasso with spike and slab priors by Xu and Ghosh (2015) \cite{xu2015bayesian}. They state that this method can produce reliable standard errors without any extra efforts. Their procedure consists of a multivariate point mass mixture similar to Zhang et al.(2014) \cite{10.1111/rssc.12053} and produces exact 0 estimates at the group level to facilitate group variable selection. The hierarchical Bayesian group lasso model with an independent spike and slab type prior for each factor $\beta_j$ is as follows:
   
\begin{equation*}
    Y|X,\beta,\sigma^2 \sim N_n(X\beta,\sigma^2I_n),
\end{equation*}
\begin{equation*}
    \beta_j|\sigma^2,\tau^2_j\overset{ind}{\sim}(1-\pi_0)N_{p_*}(0,\sigma^2\tau^2_jI_{p_*})+\pi_0\delta_0(\beta_j),\text{   } j = 1,2,...,p
\end{equation*}
\begin{equation*}
    \tau^2_j \overset{ind}{\sim} \text{ Gamma }(\frac{p_*+1}{2},\frac{\lambda^2}{2}), \text{   } j = 1,2,...,p
\end{equation*}
\begin{equation*}
    \sigma^2 \sim \text{ Inverse Gamma }(\alpha,\gamma), \sigma^2 >0
\end{equation*}
where $\delta_0(\beta_j)$ denotes a point mass at $0 \in \mathbb{R}^{p_*}, \beta_j = (\beta_{j1},...,\beta_{jp_*})^T$. An improper limiting prior is used for $\sigma^2, \pi(\sigma^2)-1/\sigma^2$.

Instead of fixing $\pi_0$ at 1/2, which is a popular choice, a conjugate beta prior is placed on it, $\pi_0 \sim \text{ Beta }(a,b)$ where $a=b=1$ which gives a prior mean $\frac{1}{2}$ and also allows a prior spread.

The value of $\lambda$ is estimated using a Monte Carlo EM algorithm (Casella, 2001\cite{10.1093/biostatistics/2.4.485}; Park and Casella, 2008\cite{doi:10.1198/016214508000000337}). The $kth$ EM update for $\lambda$ is \begin{equation*}
    \lambda^{(k)} = \sqrt{\frac{p+G}{\sum^G_{g=1}E_{\lambda^(k-1)}[\tau^2_g|Y]}}
\end{equation*}
where the posterior expectation of $\tau^2_j$ will be replaced by the sample average of $\tau^2_j$ generated in the Gibbs sampler based on $\lambda^{(k-1)}$.

\subsubsection{Uncertainty Quantification Using Posterior Medians}

To quantify uncertainty using the spike and slab prior we use posterior median thresholding. The use of these posterior median estimators in the spike and slab model instance can select and estimate variables at the same time. In \cite{xu2015bayesian} it is shown that the posterior median estimator has the oracle property for the selection and estimation of group variables under orthogonal designs, while the group lasso has a suboptimal asymptotic estimation rate when the consistency of the variable selection is achieved. We show in our simulations that the posterior median estimator of the spike and slab model correctly selects variables as well as estimates them. Previously, there has been an emphasis on using the posterior median because of its ability to be a soft thresholding estimator like the lasso but with adaptive data thresholds \cite{silverman2004}. We show in the simulation Section \ref{sec:simulation} that the spike and slab prior method is able to identify target and noise variables using these posterior medians.

%% file: 04-simulation.tex
%\newpage
\section{Numerical Experiments}\label{sec:simulation}

We supplement the simulation Section in PROLONG and add Bayesian results. We compare the performance of the univariate BLOG with Zellner's g-prior to two frequentist benchmark methods: (a) Wald test and (b) univariate mixed-effects model on simulated data. We also compare the performance of multivariate BLOG with MBGLSS to multivariate PROLONG. We focus on variable selection performance and report false positive an true positive rates of different methods. The data is simulated in the same way as the PROLONG paper so we use the Wald test and univariate mixed-effects model result from \cite{PROLONG}.

BLOG with Zellner's g-prior and $g = \sqrt{n}$ tuning parameter is able to correctly distinguish all targets from noise with a false positive rate of less than $1\%$ across all thresholds using. All targets have a decisive Bayes factor of $> 150$. BLOG using $g = $SURE minimizer as a tuning parameter is able to correctly distinguish all targets from noise with a false positive rate of less than $1\%$ across all significant thresholds (BF $>$ 50). BLOG with MBGLSS is able to correctly identify most targets and does not pick up any noise variables.

%\color{red} [SB2LP: add a short description of the main finding] \color{black}

%Uncorrelated data will be more challenging than data with correlated predictors so both univariate and multivariate scenarios will not have any real signal to assist in identifying the targets.

%These simulations do not generate exact coefficients, as we are mainly interested in the Bayesian methods correctly identifying targets and noise in comparison to the frequentist methods. The performance of the Zellner's g-prior approach will be evaluated based on selection rates for target and noise variables in comparison to frequentist Wald test and Bayes Factors based on various thresholds in order to give sensitivity and specificity rates, across 500 simulations. The performance of the Bayesian Group LASSO spike and slab prior will be evaluated based on selection rates for target and noise variables, which will be compared to the PROLONG method, and sensitivity and specificity rates, 100 simulations.
% 1) describe key hypothesis

% 2) describe in detail the DGP, i.e. what you did in your experiment

%3) if applicable, describe the performance metric

%4) findings from your experiment

\subsection{Univariate Analysis}\label{sec:sim_univ}

% 1) describe key hypothesis
We design simulations to see how accurately Zellner's g prior correctly distinguishes targets from noise compared to univariate Wald tests.

% 2) describe in detail the DGP, i.e. what you did in your experiment
The following simulation set-up captures aspects of the real data with varying levels of signal. We first simulate 100 metabolite trajectories with 20 target metabolites and 80 noise metabolites. For all scenarios, we set $N=15$, and $T=4$ to emulate the real data. 

\begin{figure}[h!]
    \centering
    \begin{subfigure}[b]{0.3\textwidth}
        \centering
        \includegraphics[height=4cm]{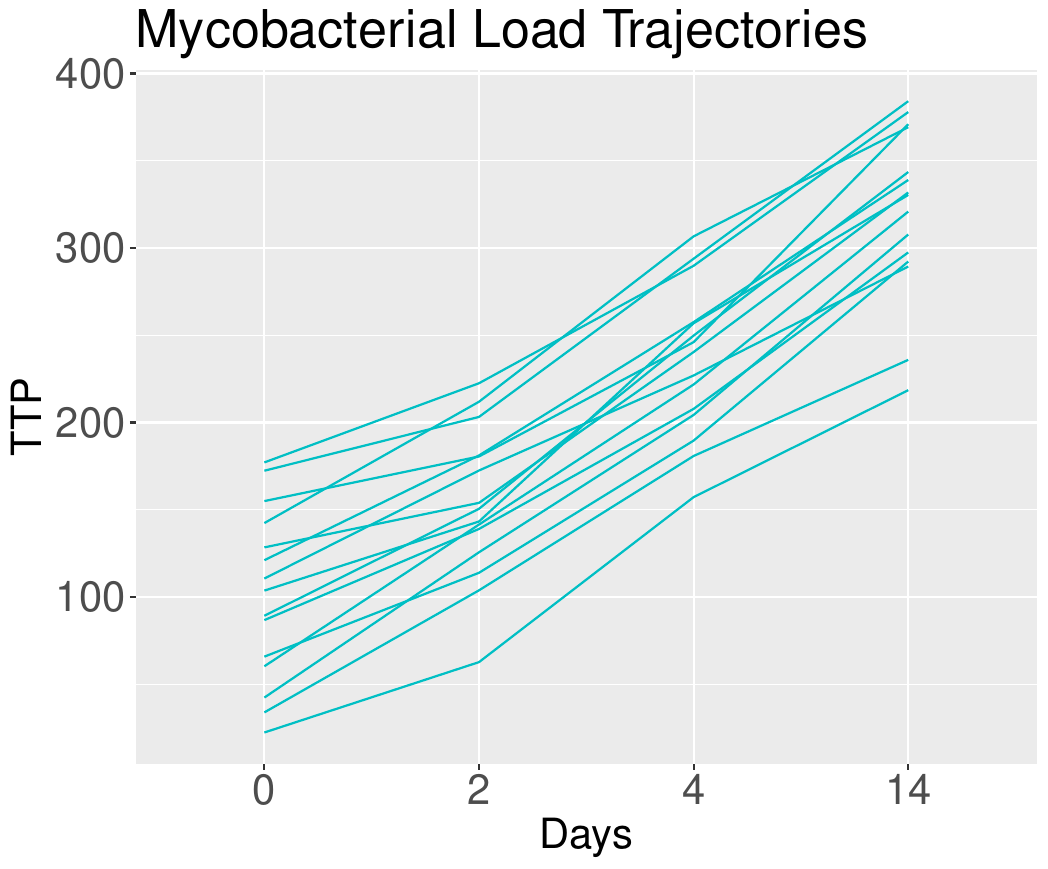}
        \caption{Y Trajectories for One Simulation}
        \label{fig:Ysim_traj}
    \end{subfigure}
    \hfill
    \begin{subfigure}[b]{0.3\textwidth}
        \centering
        \includegraphics[height=4cm]{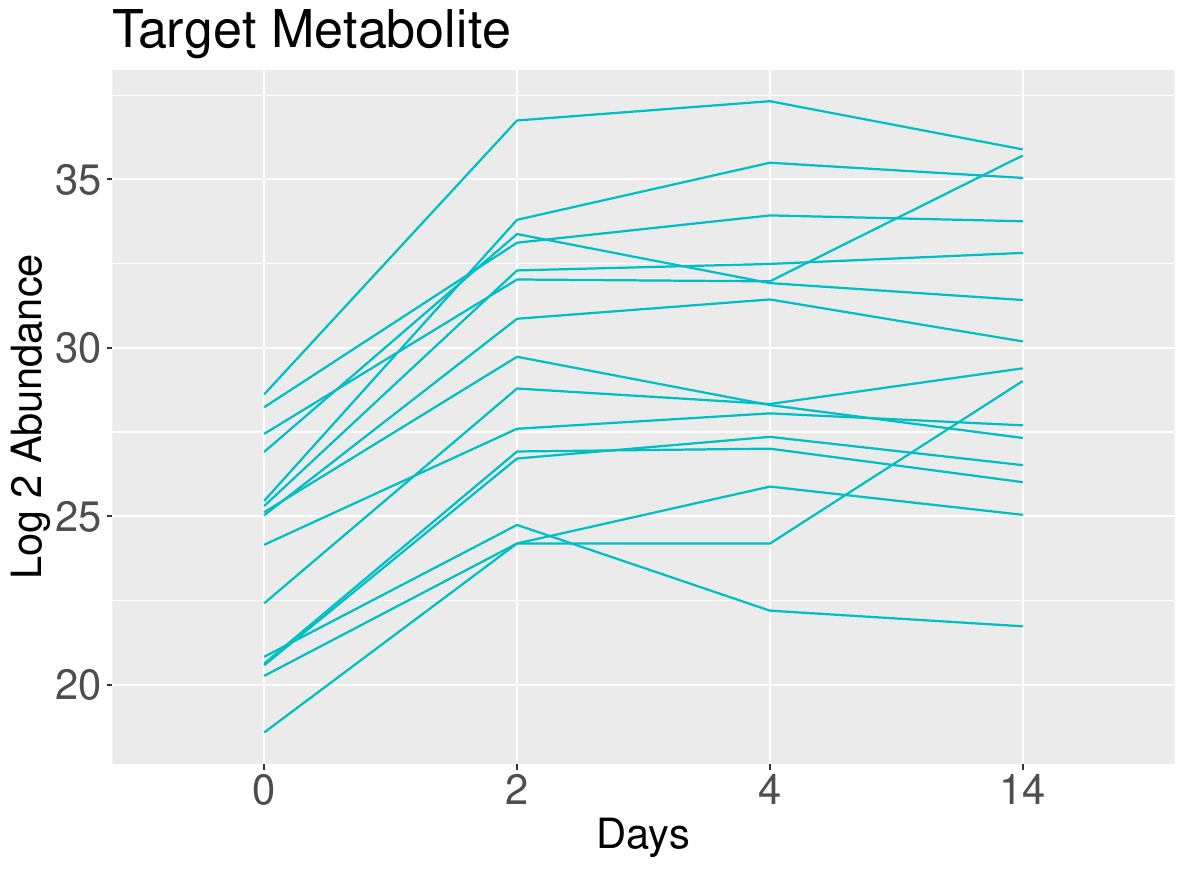}
        \caption{Example Target Metabolite Trajectory}
        \label{fig:target_sim}
    \end{subfigure}
    \hfill
    \begin{subfigure}[b]{0.3\textwidth}
        \centering
        \includegraphics[height=4cm]{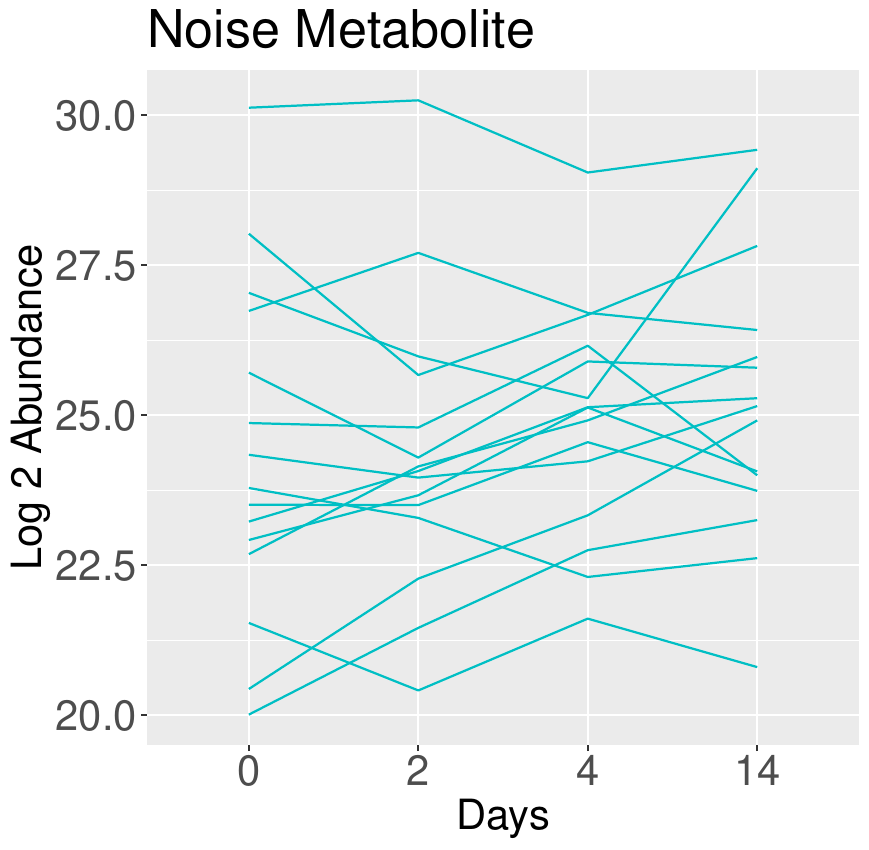}
        \caption{Example Noise Metabolite Trajectory}
        \label{fig:noise_traj}
    \end{subfigure}
    \caption{Simulation Trajectories}
    \label{fig:trajectories}
\end{figure}

In order to accurately compare the two methods, we generate data following the setup in \cite{PROLONG}. In particular, we set  $\beta = (1/3,...,1/3,0,0...,0)$ with 

\begin{equation*}
    \begin{aligned}
        y_1 &\sim N(15,5),\\
        y_2 &\sim N(y_1+\beta(x_2-x_1),5),\\
        y_3 &\sim N(y_2+\beta(x_3-x_2)+\beta(x_2-x_1),5),\\
        y_4 &\sim N(y_3+\beta(x_4-x_3)+\beta(x_3-x_2)+\beta(x_2-x_1),5),\\
        x_1 &\sim N(\mu,\Sigma_X) \text{ for } \mu \sim U(10,20),\Sigma_X = \text{ diag }(\sigma_1,...,\sigma_p), \sigma_j \sim U(1,2),\\
        x_2 &\sim x_1 + N(d\mu,\Sigma_X) \text{ for } d\mu=(5,...,10,0,...,0),\\
        x_t &\sim x_{t-1}+ N(0,\Sigma_X) \text{ for } t\in 3,4.
    \end{aligned}
\end{equation*}

We take the first differences of the $y$ vectors to mimic the same $Y$ vector setup as in equation \eqref{depvar} and follow the same $X$ matrix setup \eqref{indvar} for each metabolite as in Section \ref{sec:background}. To further test the effectiveness of Zellner's g-prior in selecting significant metabolites to noise, we simulate 30 more metabolite trajectories with 10 target and 20 noise metabolites, as well as 350 metabolite trajectories with 50 target and 300 noise metabolites.

For the Wald test, we use an $\alpha = 5\%$. For Zellner's g-prior, we consider both choices of $g$ from equation \eqref{resid} and $g = \sqrt{n}.$ Zellner's $g$ prior also requires prior coefficients to be specified. Since we are interested in whether or not a target has a nonzero coefficient, we set $p = T(T-1)/2$ prior coefficients $\beta_0=(0,...,0)$ for each of the 100 metabolites. This scenario is replicated 100 times; the results are aggregated between these replications.

\begin{figure}[htbp]
    \centering
    \begin{subfigure}[b]{0.45\textwidth}
        \centering
        \includegraphics[width=\textwidth]{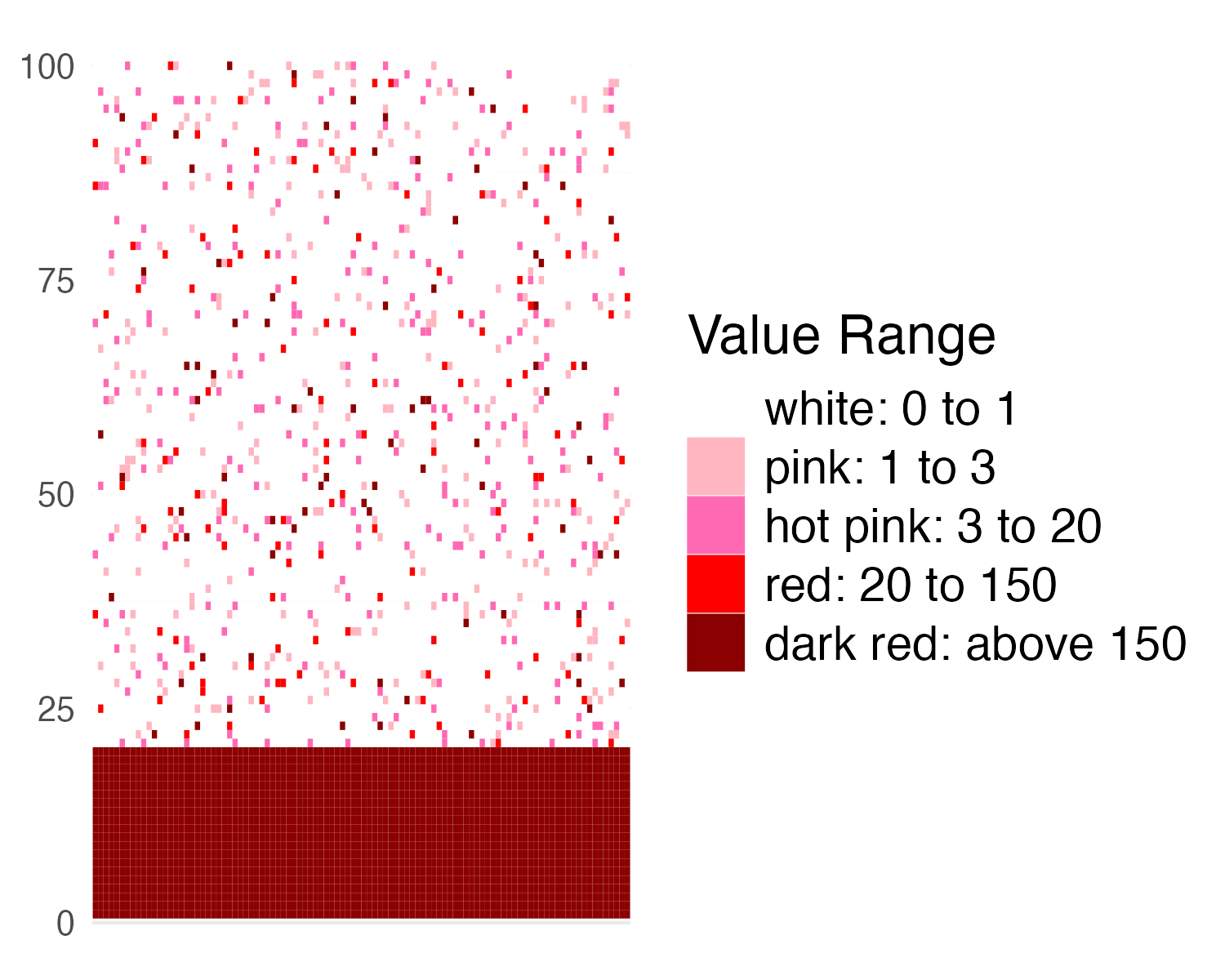}
        \caption{$g=\sqrt{n}$, 100 Variables}
        \label{fig:sqrtnsim}
    \end{subfigure}
    \hfill
    \begin{subfigure}[b]{0.45\textwidth}
        \centering
        \includegraphics[width=\textwidth]{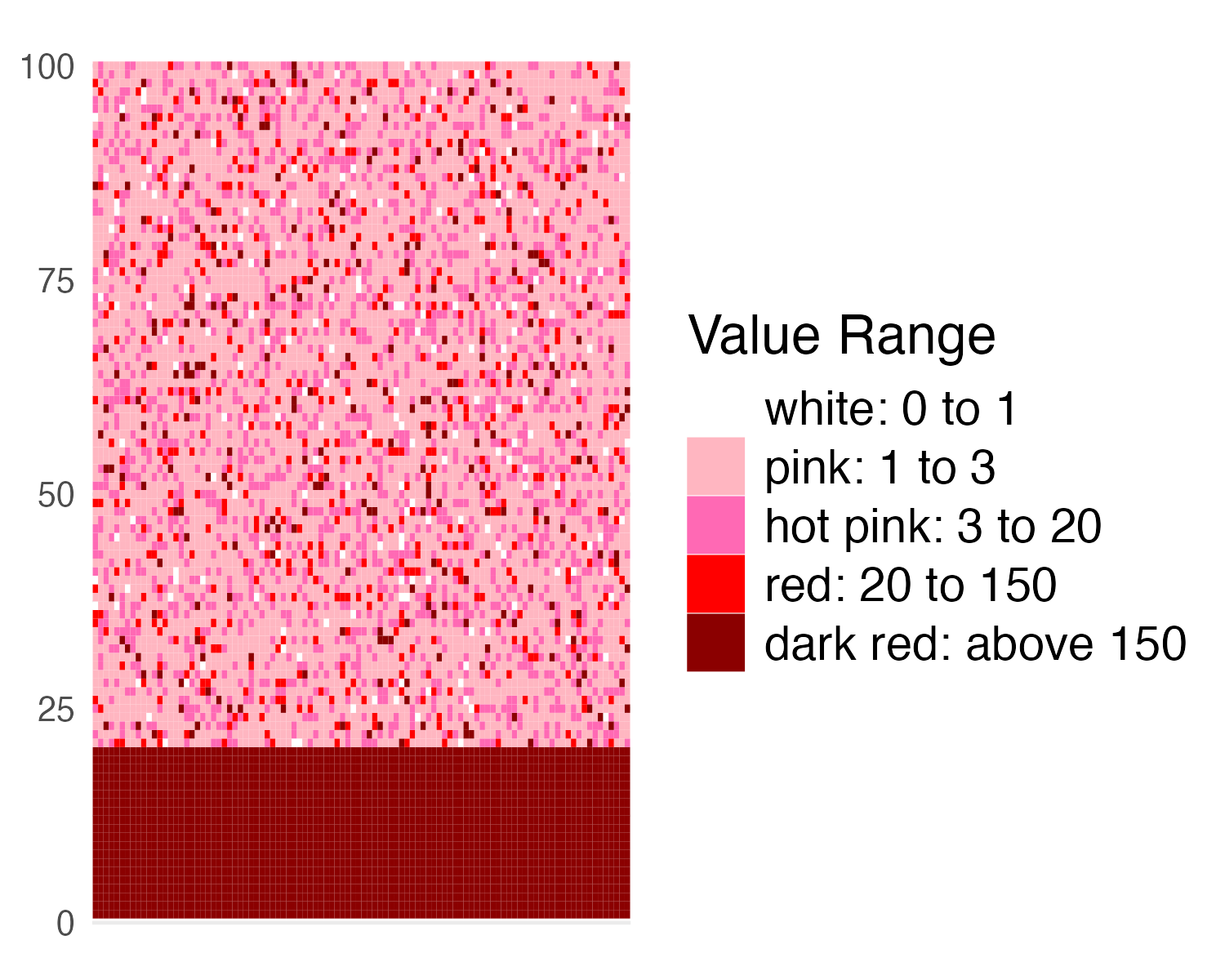}
        \caption{SURE minimizer $g$, 100 Variables}
        \label{fig:suresim}
    \end{subfigure}

    \begin{subfigure}[b]{0.45\textwidth}
        \centering
        \includegraphics[width=\textwidth]{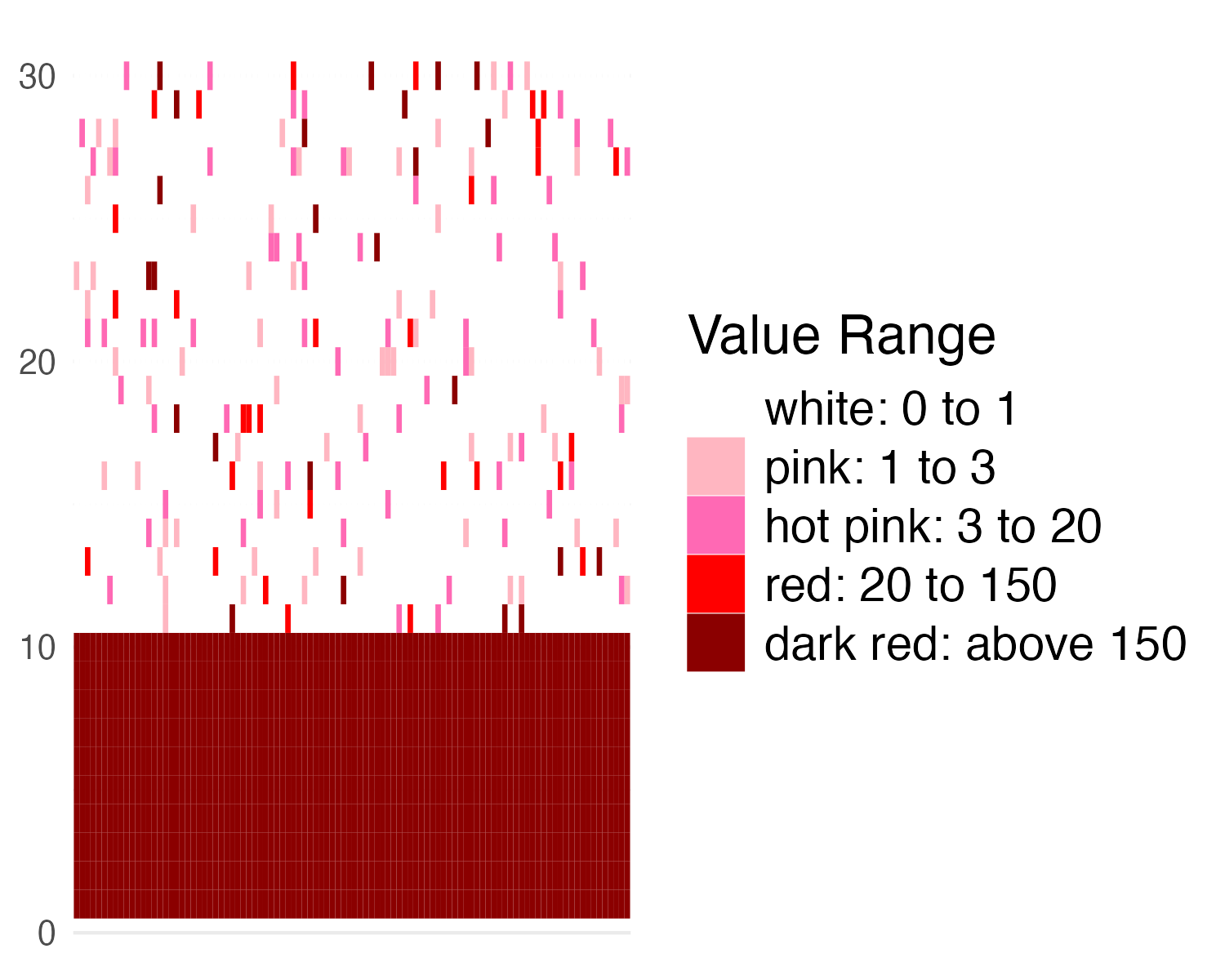}
        \caption{$g=\sqrt{n}$, 30 Variables}
        \label{fig:sqrtnsim30}
    \end{subfigure}
    \hfill
    \begin{subfigure}[b]{0.45\textwidth}
        \centering
        \includegraphics[width=\textwidth]{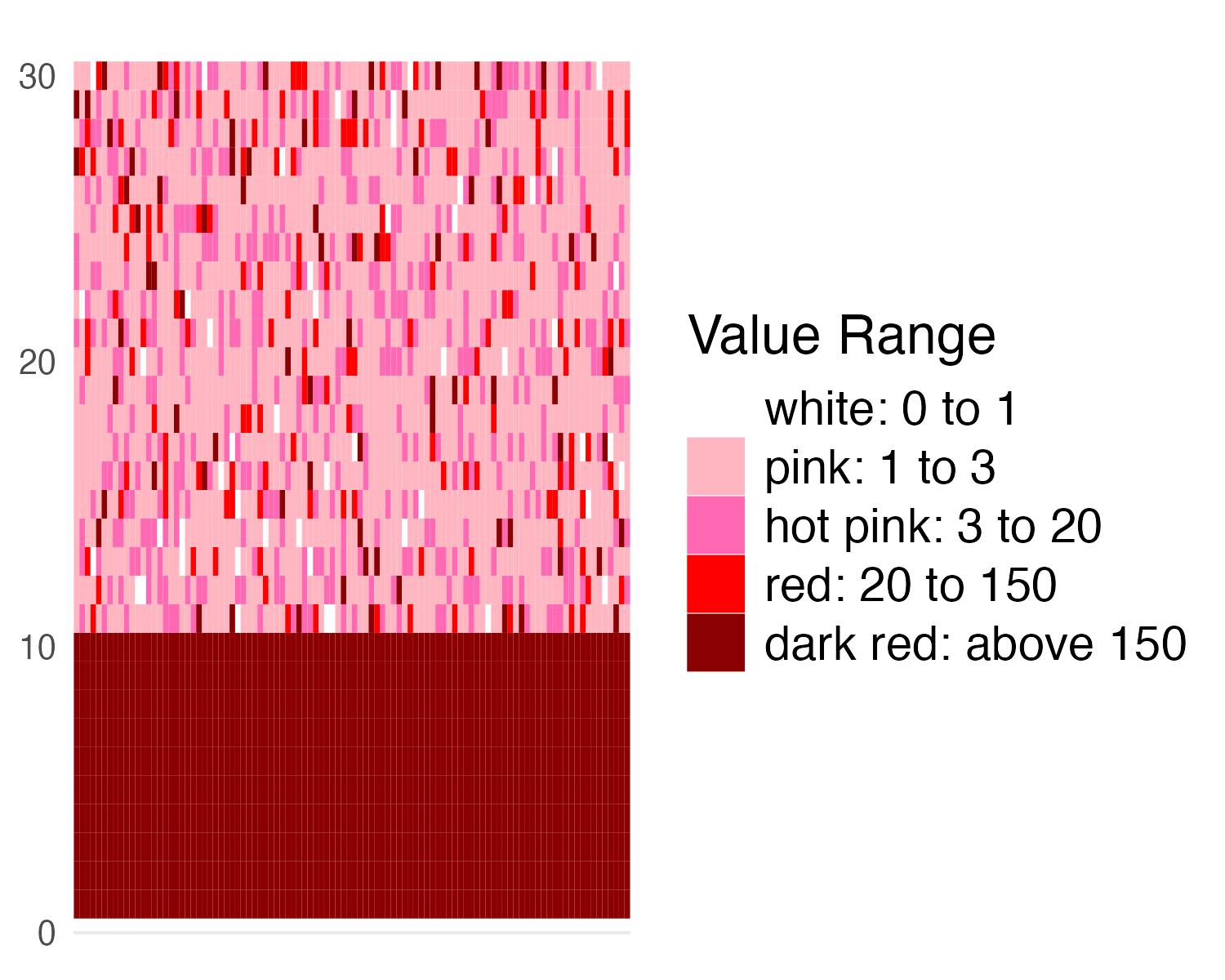}
        \caption{SURE minimizer $g$, 30 Variables}
        \label{fig:suresim30}
    \end{subfigure}

    \begin{subfigure}[b]{0.45\textwidth}
        \centering
        \includegraphics[width=\textwidth]{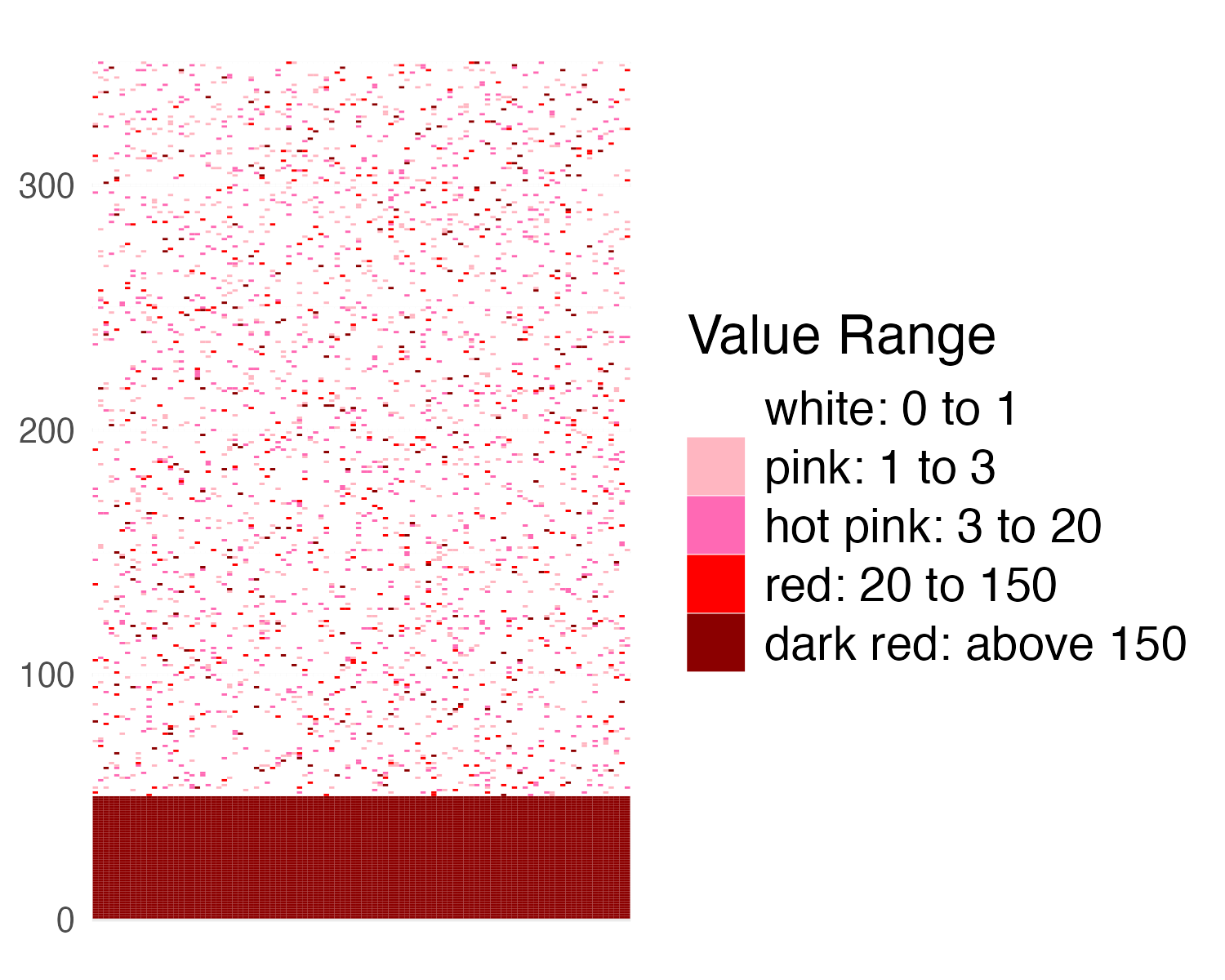}
        \caption{$g=\sqrt{n}$, 350 Variables}
        \label{fig:sqrtnsim350}
    \end{subfigure}
    \hfill
    \begin{subfigure}[b]{0.45\textwidth}
        \centering
        \includegraphics[width=\textwidth]{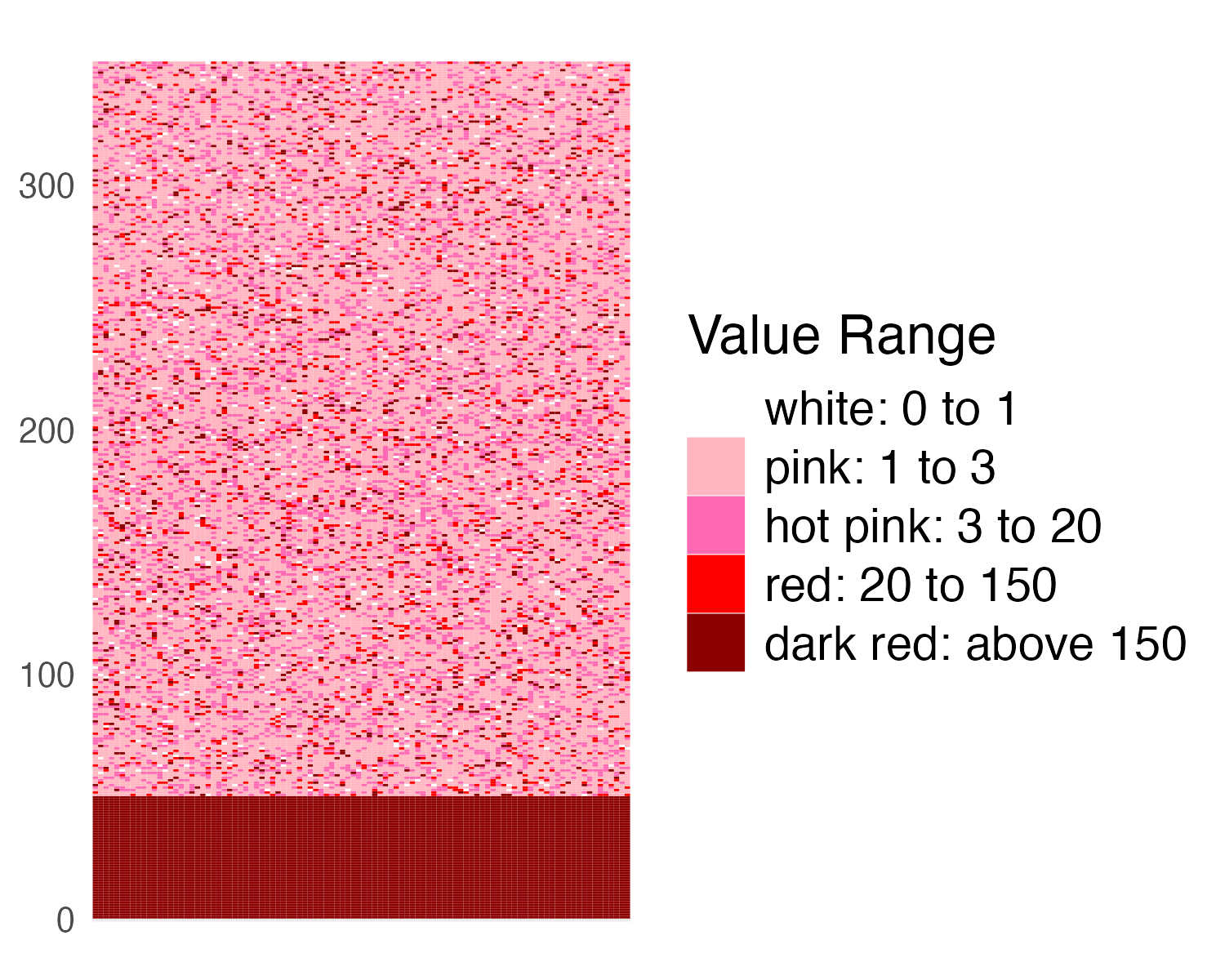}
        \caption{SURE minimizer $g$, 350 Variables}
        \label{fig:suresim350}
    \end{subfigure}

    \caption{Bayes Factor Thresholds for Zellner's $g$-prior across different simulation scenarios for various signal to noise ratios. Across all simulations, for both choices of $g$, all target variables are selected, and all have Bayes factors with very strong evidence of significance. We see that the $g = \sqrt{n}$ picks up less noise than the SURE minimizer $g$ based on the Bayes factor thresholds from table \ref{tbl:BF}.}
    \label{fig:combinedbf}
\end{figure}

\begin{figure}[htbp]
    \centering
    \begin{subfigure}[b]{\textwidth}
        \centering
        \includegraphics[width=4in]{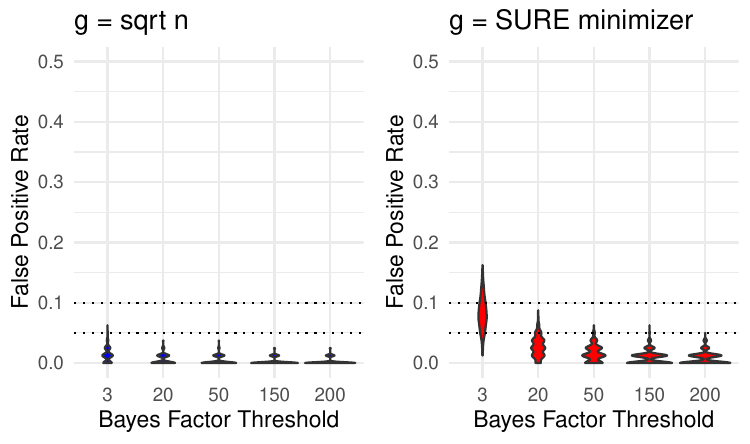}
        \caption{30 variables, 10 target 20 noise False Positive Rates}
        \label{fig:30FPrates}
    \end{subfigure}
    
    \begin{subfigure}[b]{\textwidth}
        \centering
        \includegraphics[width=4in]{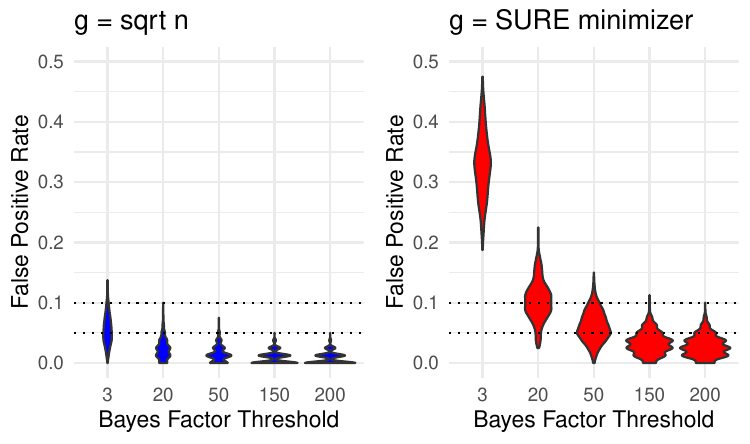}
        \caption{100 variables, 20 target 80 noise False Positive Rates}
        \label{fig:100FPrates}
    \end{subfigure}

    \begin{subfigure}[b]{\textwidth}
        \centering
        \includegraphics[width=4in]{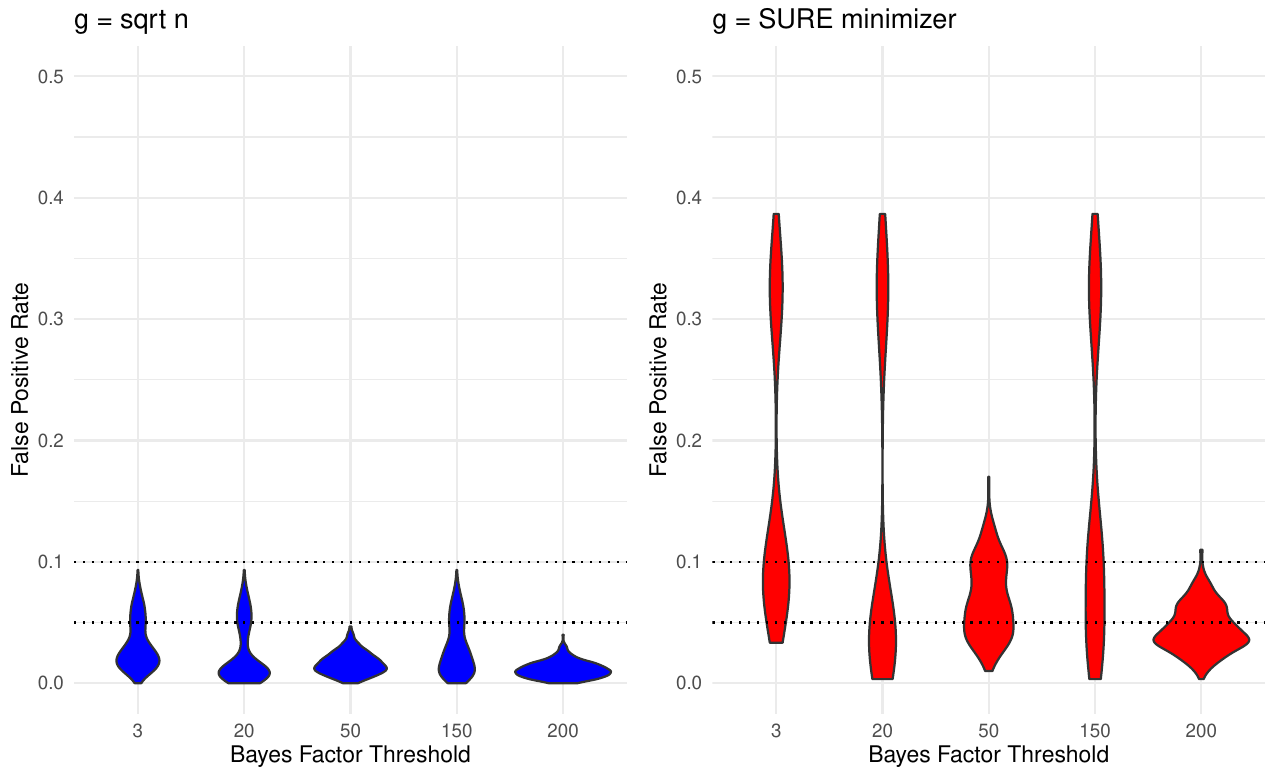}
        \caption{350 variables, 50 target 300 noise False Positive Rates}
        \label{fig:350FPrates}
    \end{subfigure}
    
    \caption{False positive rates for the 3 different variable simulation scenarios. We see that across all different simulations, $g = \sqrt{n}$ consistently has a lower false positive rate than the SURE minimizer $g$. As the signal to noise ratio decreases (i.e. more variables, slightly more targets), the false positive rate increases more significantly for the SURE minimizer $g$.}
    \label{fig:combined_FPrates}
\end{figure}

%3) if applicable, describe the performance metric
We use Bayes factor thresholds as described in Section \ref{sec: method} in Table \ref{tbl:BF}. We calculate False positive and True positive rates for both choices of $g$ for both signal-to-noise ratios.
%4) findings from your experiment
We find that for both choices of $g$, the Zellner $g$ prior method is capable of correctly identifying all 20 targets. From Figure \ref{fig:sqrtnsim} it is shown that the tuning parameter of $g = \sqrt{n}$ is more able to distinguish the targets from noise variables, as most Bayes factors are in the 0 to 1 range, meaning they are not significant. Based on the threshold of $>150$, the $g = \sqrt{n}$ tuning parameter has a false positive rate of less than 5\% across all 100 simulations as seen by Figure \ref{fig:30FPrates}. From Figure \ref{fig:suresim} we see that the tuning parameter of $g$ as the SURE minimizer is also able to distinguish the targets from noise variables, but most Bayes factors are in the 3 to 20 range. Although these noise variables have Bayes factors in the "positive" classification, based on the threshold of $>150$ the SURE minimizer tuning parameter still maintains a false positive rate of less than 10\% across all 100 simulations as seen by Figure \ref{fig:100FPrates}. The univariate model can also distinguish the 50 target variables from the 300 noise variables in the 350 variable setting, as seen by \ref{fig:suresim350} and \ref{fig:sqrtnsim350}.

This simulation performance is similar to that of PROLONG \cite{PROLONG} except now we are able to quantify uncertainty with the calculated Bayes factors.

Figure \ref{fig:30FPrates} shows the false positive rates for both choices of $g$ among 10 target variables and 20 noise variables based on different Bayes factor thresholds. We can see that for all Bayes factor thresholds, the false positive rate does not exceed 10\% and for all significant Bayes factor thresholds, the false positive rate does not surpass 5\%.

Figure \ref{fig:100FPrates} shows the false positive rates for both choices of $g$ among 20 target variables and 80 noise variables based on different Bayes factor thresholds. We can see that for significant Bayes factors, the false-positive rate does not go above 10\%. 

Figure \ref{fig:350FPrates} shows the false positive rates for both choices of $g$ among 50 target variables and 300 noise variables based on different Bayes factor thresholds. We can see that for all Bayes factors generated using $g=\sqrt{n}$, the false positive rate is below 10\%.

\subsection{Multivariate Analysis}\label{sec:sim_multiv}

% 1) describe key hypothesis
The objectives of the multivariate simulations aim to compare the performance of the Bayesian group lasso with the spike and slab prior method to PROLONG.
% 2) describe in detail the DGP, i.e. what you did in your experiment

The simulation set-up is the same matrix set-up as presented in Section \ref{sec:MVM}. We simulate 3 different variable scenarios to see how the multivariate method performs on different size datasets. We simulate 10 targets with 20 noise variables, 20 targets with 80 noise variables, and 50 targets with 300 noise variables for 100 simulations each. 

%3) if applicable, describe the performance metric
We evaluated the performance of the Bayesian method by comparing the selection counts of the target variables with those of PROLONG. We are also able to give an uncertainty quantification metric by providing nonzero posterior medians for the target variables, as described in Section \ref{sec: method}.

%4) findings from your experiment
We present the zero and nonzero coefficients and medians for each variable scenario below. We see that the multivariate Bayesian method is able to identify target coefficients and also provide nonzero medians.

\begin{figure}[htbp]
    \centering
    % First row: 30 variables
    \begin{subfigure}[b]{0.48\textwidth}
        \centering
        \includegraphics[width=2.78in]{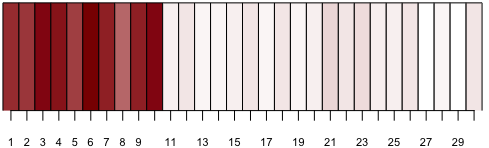}
        \caption{Posterior Medians (30 variables)}
        \label{fig:30pos_medians}
    \end{subfigure}
    \hfill
    \begin{subfigure}[b]{0.50\textwidth}
        \centering
        \includegraphics[width=2.7in, height = 0.84in]{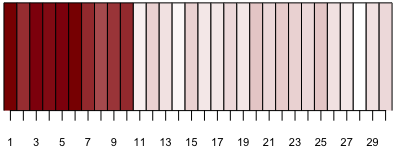}
        \caption{Coefficients (30 variables)}
        \label{fig:30gllcount}
    \end{subfigure}
    
    % Space between rows
    \vskip\baselineskip
    
    % Second row: 100 variables
    \begin{subfigure}[b]{0.48\textwidth}
        \centering
        \includegraphics[width=2.7in, height = 0.84in]{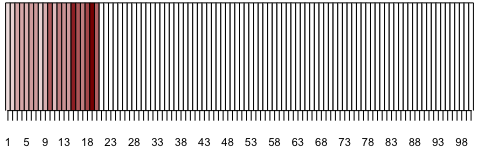}
        \caption{Posterior Medians (100 variables)}
        \label{fig:100pos_medians}
    \end{subfigure}
    \hfill
    \begin{subfigure}[b]{0.48\textwidth}
        \centering
        \includegraphics[width=2.7in, height = 0.84in]{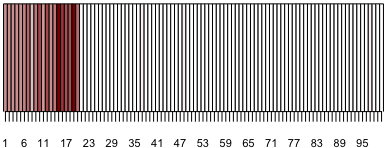}
        \caption{Coefficients (100 variables)}
        \label{fig:100gllcount}
    \end{subfigure}
    
    % Space between rows
    \vskip\baselineskip
    
    % Third row: 350 variables
    \begin{subfigure}[b]{0.48\textwidth}
        \centering
        \includegraphics[width=2.75in, height = 0.81in]{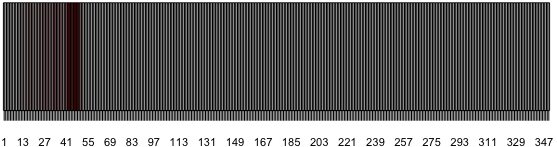}
        \caption{Posterior Medians (350 variables)}
        \label{fig:350pos_medians}
    \end{subfigure}
    \hfill
    \begin{subfigure}[b]{0.48\textwidth}
        \centering
        \includegraphics[width=2.7in, height = 0.84in]{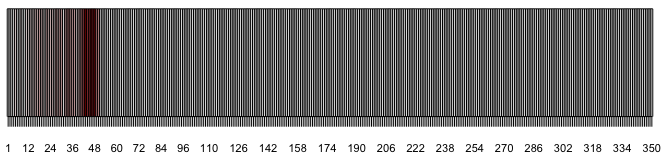}
        \caption{Coefficients (350 variables)}
        \label{fig:350gllcount}
    \end{subfigure}

    \caption{Posterior Medians (left column) and Coefficients (right column) for 3 different simulation scenarios. We see that none of the noise variables are picked up for the 20 target 80 noise and 50 target 300 noise. Not all target coefficients are identified and there are some zero posterior medians for these signal to noise ratios. For the 10 target, 20 noise simulation the multivariate BLOG method is able to distinguish target from noise the most distinctly. All target coefficients are identified along with nonzero posterior medians, but some noise variables are picked up throughout few simulations in this scenario.}
    \label{fig:3x2_grid}
\end{figure}

Figure \ref{fig:3x2_grid} shows that for the target variables, the coefficients are non-zero and are significantly different from the noise variables.

The BLOG model does not identify any noise variables as a target, but is also unable to identify all target variables for the 350 variable simulation. This could be due to the small signal-to-noise ratio causing the model to place more weight on the noise.

\begin{figure}[h!]
    \centering
    \includegraphics[width=3in]{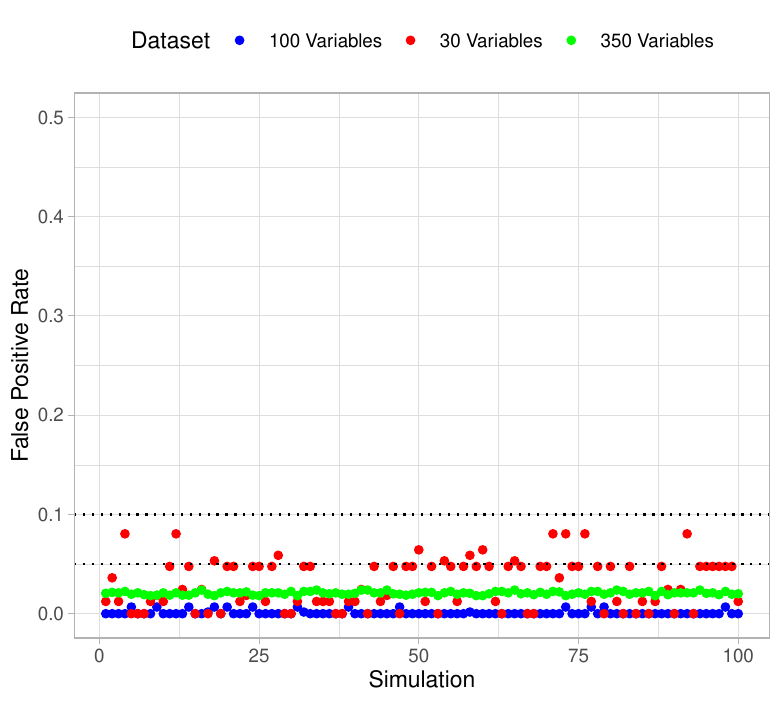}
    \caption{False positive rate across simulations for 30, 100, and 350 variables. We see that for all simulation scenarios the false positive rate is under 10\% and for 100 and 350 variables the false positive rate is under 5\%.}
    \label{fig:fpratesmedians}
\end{figure}

We also present false positive rates for all simulation scenarios in \ref{fig:fpratesmedians}. The average false positive rate across 100 simulations for 30 variables was 0.03, for 100 variables was 0.0009 and for 350 variables it was 0.021.

%Below is a table comparing some target and noise coefficients and their medians. The table shows that the posterior medians are zero for most noise variables and nonzero for the targets, meaning the model is correctly distinguishing the targets from the noise variables. 

%% file: 05-realdata.tex
%\newpage
\section{Real Data Analysis}\label{sec:realdata}
We illustrate the benefit of using BLOG instead of existing benchmarks on two real clinical data sets on TB and diabetes.

\subsection{TB Data Analysis}
%\subsection{Clinical Significance}
Tuberculosis (TB) is a preventable and usually curable disease. However, in 2022, tuberculosis was the second leading cause of death from a single infectious agent, after coronavirus disease (COVID-19), and caused almost twice as many deaths as HIV/AIDS \cite{who2023}. A new shortened treatment regimen can have potential advantages, including improving outcomes through increased adherence, reduced cure time, reducing costs incurred by patients, and reducing treatment delivery costs incurred by health systems \cite{gomez2016cost}. This study was a clinical trial to determine whether an FDA-approved antiparasitic drug could work as a bacterial killing treatment against Mycobacterium tuberculosis. Sputum samples were collected every two days to monitor changes in culture-based time to positivity (TTP), which is inversely correlated with bacterial load. Additionally, urine samples were collected before treatment on day 0, and on days 2, 4, and 14 post-treatment and the abundance of urinary metabolites were measured via metabolomic profiling. The PROLONG article provides more reasoning for the use of this particular dataset, stating that these data are reflective of a common collection of biomarkers at a few time points with a continuous outcome of interest, while also providing a challenge that existing methods are not intended to address. Since we are using the same dataset as was used in PROLONG's \cite{PROLONG} data analysis, we are able to quantify uncertainty which address the reliability of predictions and will further assist in decision-making from these Bayesian methods.

%\subsection{Collection and Preprocessing}
We used the same TB omics data that are used in the PROLONG \cite{PROLONG} article to maintain an accurate comparison of our Bayesian method to PROLONG. In the article, they describe the imputation method and the data limiting criterion they used to obtain the 4 time-point matrices each containing 352 metabolite abundances for each subject. 

%\subsection{Benchmark Method}
We used the selected metabolites for the PROLONG, Wald tests, and univariate longitudinal mixed-effect models that were targets identified in \cite{PROLONG}. We compare the Bayes factors for each metabolite produced by the Bayesian univariate approach using Zellner's g-prior to the Wald tests and univariate longitudinal mixed models. We also compare the credible intervals produced by the multivariate Bayesian group lasso with the spike and slab prior to PROLONG. 

%\subsection{BLOG Results}

Table \ref{table:unirealdata} below shows the top 20 metabolites ordered by the lowest Wald test p-value. The Wald test is known to pick up extra noise, meaning that the metabolites are overly selected. Ideally, we want our BLOG method to select a number of metabolites that are between the Wald test and the LME model. Both choices of $g$ select almost all metabolites found by the Wald test and all have Bayes factors greater than 150. This is of similar performance to PROLONG but now we have uncertainty quantification through Bayes factors.

\begin{figure}[ht]
    \centering
    \includegraphics[width=3.5in]{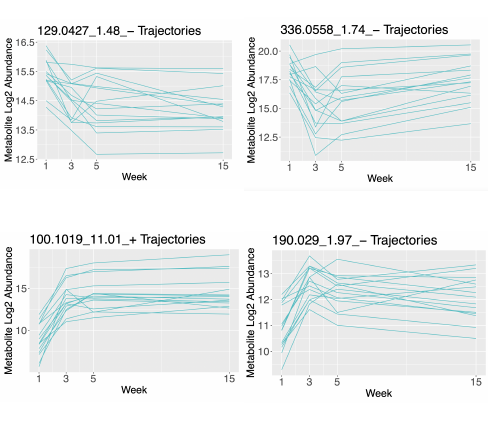}
    \caption{Trajectories of significant metabolites identified by univariate BLOG}
    \label{fig:significantmetabs}
\end{figure}

When using the tuning parameter of $g = \sqrt{n}$, Zellner's g prior approach gives 38 metabolites with Bayes factors $>150$, with the metabolites with the top 6 highest values being 204.1867\_9.3\_+, 100.1019\_11.01\_+, 336.0558\_1.74\_-, 129.0427\_1.48\_-, 284.1505\_1.69\_+, and 344.0923\_0.81\_-. The SURE minimizer choice of $g$ also selects 38 metabolites with Bayes factors $>150$. The metabolites with the highest six values in the order are the same as the $g=\sqrt{n}$ choice. Both tuning parameters produce metabolites in a similar order as the top 6 metabolites with the lowest p values using the Wald test, as seen in Table \ref{table:unirealdata}. We also plot some examples of significant metabolite trajectories that the univariate model picks up in Figure \ref{fig:significantmetabs}. This shows that the model can pick up predictors that show an increase and a decrease over time. This is of interest to clinical investigators because they may not initially pick these metabolites to be significant, which could provide new insights into biomarker identification.

\begin{table}[ht]
\caption{Top 20 Metabolites by Lowest P-Values with Bayes Factors Checkmarks} 
\centering
\begin{tabular}{lllll}
  \hline
Metabolite & Wald Test p-value & Bayes Factor $\sqrt{n}$  & Bayes Factor Sure & LME p-value \\ 
  \hline
204.1867\_9.3\_+ & \checkmark & \checkmark & \checkmark &  \\ 
  129.0427\_1.48\_- & \checkmark & \checkmark & \checkmark &  \\ 
  336.0558\_1.74\_- & \checkmark & \checkmark & \checkmark &  \\ 
  100.1019\_11.01\_+ & \checkmark & \checkmark & \checkmark &  \\ 
  344.0923\_0.81\_- & \checkmark & \checkmark & \checkmark & \checkmark \\ 
  168.0282\_1.74\_- & \checkmark & \checkmark & \checkmark &  \\ 
  284.1505\_1.69\_+ & \checkmark & \checkmark & \checkmark &  \\ 
  183.0531\_2\_- & \checkmark & \checkmark & \checkmark &  \\ 
  429.058\_1.7\_- & \checkmark & \checkmark & \checkmark &  \\ 
  190.029\_1.97\_- & \checkmark & \checkmark & \checkmark &  \\ 
  406.1291\_2.28\_- & \checkmark & \checkmark & \checkmark & \checkmark \\ 
  246.0822\_6.62\_+ & \checkmark & \checkmark & \checkmark & \checkmark \\ 
  304.0643\_1.63\_- & \checkmark &  &  &  \\ 
  247.1558\_1.56\_+ & \checkmark & \checkmark & \checkmark &  \\ 
  241.0543\_2.07\_- & \checkmark &  &  &  \\ 
  440.1626\_1.11\_- & \checkmark &  &  &  \\ 
  267.095\_7.78\_- & \checkmark & \checkmark & \checkmark & \checkmark \\ 
  362.0698\_1.78\_- & \checkmark &  &  &  \\ 
  281.1882\_10.95\_+ & \checkmark & \checkmark & \checkmark & \checkmark \\ 
  322.0567\_2.13\_- & \checkmark &  &  & \checkmark \\ 
   \hline
\end{tabular}
\label{table:unirealdata}
\end{table}

\subsection{Diabetes Data Analysis}

%\subsection{Clinical Significance}

Type 2 diabetes affects people of increasingly younger age. Current recommendations for diabetes management focus strongly on the use of medications to control blood glucose, blood lipids, and blood pressure, but this results in substantial patient burden and healthcare costs. Instead, this clinical study aims to focus on addressing the underlying reversible causes of diabetes. This study is the first randomized controlled trial in young patients with early diabetes ($\leq 3$ years) in primary care who are from the Middle East and North Africa region that examines the effect of a 12-month intensive lifestyle intervention.
\cite{taheri2020diadem} This involves the incorporation of a total diet replacement phase, focusing mainly on weight loss and glycemic control. This data set is of similar structure to the previous TB omics data, so we can further explore the ability of BLOG to quantify uncertainty. We use the method in 3 outcomes, where we aim to see how the proteins vary in time with the outcomes of HDL (high-density lipoprotein), I2GDF15 (fat protein) and I2MIC1 (fat protein) in 16 subjects.

%\subsection{Collection and Preprocessing}

We used data on diabetes collected from a randomized controlled trial conducted in primary care and community settings in Qatar that compared the effects of an intensive lifestyle intervention with usual medical care on weight loss and glycemic outcomes in individuals with type 2 diabetes. Data is collected over 12 months in five time-points but the last time-point is removed due to lack of data. We only test BLOG on 4 time points each containing 17 protein abundances that were of interest for each subject. 

%\subsection{BLOG Results}

Tables \ref{tab:ILIBFHDL}, \ref{tab:ILIBFI2GDF}, \ref{tab:ILIBFI2MIC1} show Bayes factors for each protein based on each outcome. We see that the Bayes factors for the treatment group that is treated with an intensive lifestyle intervention (ILI) are significantly higher than that of the control group that received Basic Medical Care (BMC) for the significant proteins across all three outcomes. 

The multivariate method was also applied to this dataset and for the ILI treatment with the I2GDF15 outcome and found a nonzero median for the l2ApoD protein. This one has a semi-significant Bayes factor from the univariate results. The method also found that from the ILI treatment with the I2MIC1 outcome the l2HeparincofactorII protein had nonzero medians. This means that our BLOG method is able to identify proteins co-varying over time that can lead to an increase or decrease in fat and cholesterol levels, giving insight into exactly which biomarkers are significant when trying to identify the cause of a reversal of type 2 diabetes.

\begin{table}
\caption{ILI Treatment Bayes Factors for HDL}
\centering
\resizebox{0.9\linewidth}{!}{
\begin{tabular}[t]{lrr}
\toprule
Protein & Bayes Factor sqrtn & Bayes Factor Sure\\
\midrule
l2ApoE4\_ & 21.809 & 35.578\\
l2ApoE\_ & 15.177 & 26.602\\
l2ApoL1\_ & 15.177 & 26.602\\
l2ApoB\_ & 8.184 & 16.450\\
l2ApoptosisregulatorBclW\_ & 7.476 & 15.356\\
l2ApoM\_ & 2.057 & 6.066\\
l2ApoD\_ & 1.515 & 4.945\\
l2ApoAI\_ & 1.261 & 4.392\\
l2ApoE2\_ & 0.464 & 2.419\\
l2complementfactorHrelated5\_ & 0.352 & 2.086\\
l2Hemopexin\_ & 0.276 & 1.848\\
Clusterin\_ & 0.106 & 1.245\\
l2HeparincofactorII\_ & 0.084 & 1.157\\
l2KininogenHMW\_ & 0.076 & 1.126\\
l2amyloidprecursorprotein\_ & 0.067 & 1.093\\
l2GSK3alphabeta\_ & 0.061 & 1.070\\
l2paraoxonase1\_ & 0.006 & 2.430\\
\bottomrule
\end{tabular}}
\label{tab:ILIBFHDL}
\end{table}

\begin{table}
\caption{ILI Treatment Bayes Factors for I2GDF15}
\centering
\resizebox{0.9\linewidth}{!}{
\begin{tabular}[t]{lrr}
\toprule
Protein & Bayes Factor sqrtn & Bayes Factor Sure\\
\midrule
l2ApoE4\_ & 3662.265 & 3671.162\\
l2ApoE\_ & 366.490 & 409.976\\
l2ApoL1\_ & 231.234 & 269.733\\
l2ApoB\_ & 25.332 & 40.189\\
l2ApoptosisregulatorBclW\_ & 25.332 & 40.189\\
l2ApoM\_ & 10.328 & 19.672\\
l2ApoD\_ & 2.003 & 5.956\\
l2ApoAI\_ & 0.822 & 3.364\\
l2ApoE2\_ & 0.321 & 1.991\\
l2complementfactorHrelated5\_ & 0.277 & 1.853\\
l2Hemopexin\_ & 0.220 & 1.661\\
Clusterin\_ & 0.108 & 1.252\\
l2HeparincofactorII\_ & 0.047 & 1.026\\
l2KininogenHMW\_ & 0.030 & 1.000\\
l2amyloidprecursorprotein\_ & 0.018 & 1.089\\
l2GSK3alphabeta\_ & 0.014 & 1.206\\
l2paraoxonase1\_ & 0.011 & 1.378\\
\bottomrule
\end{tabular}}
\label{tab:ILIBFI2GDF}
\end{table}

\begin{table}
\caption{ILI Treatment Bayes Factors for I2MIC1}
\centering
\resizebox{0.9\linewidth}{!}{
\begin{tabular}[t]{lrr}
\toprule
Protein & Bayes Factor sqrtn & Bayes Factor Sure\\
\midrule
l2ApoE4\_ & 82.039 & 108.006\\
l2ApoE\_ & 2.648 & 7.214\\
l2ApoL1\_ & 1.074 & 3.966\\
l2ApoB\_ & 0.625 & 2.860\\
l2ApoptosisregulatorBclW\_ & 0.607 & 2.811\\
l2ApoM\_ & 0.607 & 2.811\\
l2ApoD\_ & 0.249 & 1.759\\
l2ApoAI\_ & 0.201 & 1.596\\
l2ApoE2\_ & 0.085 & 1.162\\
l2complementfactorHrelated5\_ & 0.060 & 1.069\\
l2Hemopexin\_ & 0.019 & 1.059\\
Clusterin\_ & 0.013 & 1.233\\
l2HeparincofactorII\_ & 0.009 & 1.682\\
l2KininogenHMW\_ & 0.007 & 2.291\\
l2amyloidprecursorprotein\_ & 0.006 & 2.385\\
l2GSK3alphabeta\_ & 0.005 & 3.922\\
l2paraoxonase1\_ & 0.004 & 8.971\\
\bottomrule
\end{tabular}}
\label{tab:ILIBFI2MIC1}
\end{table}

%% file: 06-conclusion.tex
%\newpage
\section{Conclusion}\label{sec:conclusion}
In this work, we introduced a Bayesian framework for high-dimensional regression and variable selection in longitudinal omics datasets, addressing the growing need for uncertainty quantification in biomarker discovery. We developed and validated two methodologies: the univariate Bayesian approach using Zellner's $g$ prior and Bayes factors, and the multivariate Bayesian group lasso with spike and slab priors. The univariate approach improves traditional regression methods by providing Bayes factors to quantify evidence for variable inclusion, while the Bayesian group lasso framework extends variable selection to group-level interactions, incorporating correlation structures across metabolites over time. Both approaches are designed to handle the complexities of longitudinal omics data, where the interplay between time-varying outcomes and large sets of predictors is critical.

Throughout this article, we have shown the validity of BLOG as a way to quantify uncertainty through the figures and tables presented. The univariate version is capable of correctly identifying target variables in simulations and providing significant Bayes factors that quantify uncertainty. In the multivariate case, BLOG is able to identify target variables by giving nonzero posterior medians in various simulation scenarios. The false positive rates in simulations for both univariate and multivariate methods are lower than 10\% when identifying the target variables. 

We applied our methods to two real-world datasets: a tuberculosis study and an intensive lifestyle intervention study aimed at understanding diabetes. Our univariate Bayesian framework successfully identified significant biomarkers in both studies. Compared to the univariate baseline PROLONG method, the univariate method selects very similar metabolites in the real data scenario. Identification of significant biomarkers demonstrates the ability of BLOG to uncover biologically significant patterns and provide valuable information on disease progression and response to treatment.

These contributions are novel in the sense that this uncertainty quantification model has not been applied to these types of longitudinal omics datasets. The use of a Bayesian framework in the discovery of biomarkers provides a new perspective on metabolite-drug and protein-outcome interactions over time. The Bayesian group lasso leverages sparsity-inducing properties and when combined with the first differencing of the data and matrix setup, facilitates the identification of target variables and uncertainty quantification. This approach can improve the predictive accuracy and improve the interpretability of the results, which can significantly help clinical researchers.

Future research directions could build on these findings by expanding BLOG to accommodate other types of omics data or exploring applications in other areas using longitudinal datasets, such as econometrics, among others.

\section*{Acknowledgments}

All authors acknowledge partial support from the NIH R01GM135926 award. S. Basu acknowledges partial support from NSF awards DMS-1812128, DMS-2210675, CAREER DMS-2239102, and NIH award R21NS120227.

%% file: 07-appendix.tex
\appendix
\setcounter{section}{0}
\renewcommand{\thesection}{\Alph{section}}
\setcounter{equation}{0}
\renewcommand{\theequation}{\Alph{section}.\arabic{equation}}
\section{Appendix}\label{sec:appendix}

\subsection{Proof of the Equality of the Slope Coefficients in the Level and Difference Models}
\begin{proof}
    For a single metabolite,  the level model in (\ref{model1}) can be expressed as
\begin{equation} \label{model_A1}
    Y_{T \times 1} = \mathbbm{1}_{T \times 1} \beta_0 + X_{T \times k} \beta_1 + \epsilon
\end{equation}
with rank($\mathbbm{1}: X) = k+1$. The difference model, analogous to (\ref{depvar}) and (\ref{indvar}), is then 
\begin{equation} \label{model_A2}
        DY = DX\beta_1 + D\epsilon 
\end{equation} 
where  $D$ is the $(T-1) \times T$ differencing matrix. 

The generalized least squares (GLS) estimate for $\beta_0$ and $\beta_1$ in (\ref{model_A1}) are \begin{equation} \label{GLS_A1}
        \begin{pmatrix}
            \hat{\beta_0} \\ \hat{\beta_1} 
        \end{pmatrix} = \begin{bmatrix}
            \mathbbm{1}^\top \Sigma^{-1} \mathbbm{1} & \mathbbm{1} \Sigma X \\
            X^\top \sigma^{-1} \mathbbm{1} & X^\top \Sigma^{-1} X
            \end{bmatrix} \begin{bmatrix}
                \mathbbm{1}^\top \sigma^{-1} Y \\
                X^\top \Sigma^{-1}Y
            \end{bmatrix}.
\end{equation}
Note that $\hat{\beta_1} = (X^\top \Omega X)^{-1}X^\top \Omega Y$ where $\Omega = \Sigma^{-1} - \Sigma^{-1} \mathbbm{1} (\mathbbm{1}^\top \Sigma \mathbbm{1})^{-1} \mathbbm{1} \Sigma^{-1}$. 

Since $Cov(D \epsilon) = D\Sigma D^\top$  the GLS estimate of $\beta_1$ under (\ref{model_A2}) is 
\begin{equation} \label{GLS_A2}
        \hat{\beta_1} = (X^\top\Psi X)^{-1} X^\top \Psi Y
\end{equation} 
where $\Psi = D^\top(D\Sigma D^\top)^{-1} D$.

Therefore, we can show that the estimates in $\beta_1$ under (\ref{GLS_A1}) and (\ref{GLS_A2}) for the models in (\ref{model_A1}) and (\ref{model_A2}), respectively, are equivalent if and only if $\Omega = \Psi$. In this case, 
\begin{equation*}
    \Sigma^{-1}-\Sigma^{-1}\mathbbm{1}(\mathbbm{1}^\top \Sigma^{-1} \mathbbm{1})^{-1}\mathbbm{1}^\top \Sigma^{-1} 
    = D^\top (D\Sigma D^\top)^{-1} D.
    \end{equation*}
This then implies that
\begin{equation*}
    I = \Sigma^{-1} \mathbbm{1} (\mathbbm{1}^\top  \Sigma^{-1} \mathbbm{1})^{-1} \mathbbm{1} + D^\top (D \Sigma D^\top )^{-1}D \Sigma    
    \end{equation*}
Note, the first term is the $\Sigma$ projection onto the $1-$dimension column space of $\Sigma^{-1}\mathbbm{1}$. Furthermore, the second term is the projection $\Sigma$ onto the $(T-1)-$ dimensional column space of $D^\top$. It then follows that these two column spaces are $\Sigma$ orthogonal complements, since $D \Sigma (\Sigma^{-1} \mathbbm{1}) = D \mathbbm{1} = 0$.
\end{proof}

\subsection{Zellner's $g$-prior for $p > n$} \label{app:zellgpgreatern}

Maruyama and George (2011)\cite{maruyama2011fully} proposed a selection criterion based on a fully Bayes formulation with a generalization of Zellner's $g$-prior which allows for $p > n$. They obtain their proposed $g$-prior Bayes factor (gBF), of the form (omitting $\gamma$ subscripts for clarity) \begin{equation*}
gBF_{\gamma:N} =  
\begin{cases}
    \left\{ \frac{\bar{d}}{d_q} \right\} 
    \frac{\{1-R^2 + d^2_q || \hat{\beta}_{LS} ||^2\}^{-1/4 - p/2}}{C_{n,q}(1 - R^2)^{(n-q)/2 - 3/4}}
    & \text{if } q < n - 1 \\
    \left\{ \bar{d} \times || \hat{\beta}^{MP}_{LS} || \right\}^{-n + 1} 
    & \text{if } q \geq n - 1
\end{cases}
\end{equation*}
where $C_{n, q} \equiv \frac{B(1/4,(n-q)/2-3/4)}{B(q/2+1/4,(n-q)/2-3/4}$ using the Beta function $B(\cdot,\cdot), R^2$ is the R-squared statistic under $\mathcal{M}_\gamma, \bar{d}$ and $d_r$ are the geometric mean and minimum of the singular values of $X_\gamma$, $||\cdot|| $ is the $L_2$ norm, $\hat{\beta}_{LS}$ is the least squares estimator, and $\hat{\beta}_{LS}^{MP}$ is the least squares estimator using the Moore-Penrose inverse matrix. Furthermore, they define 
\begin{equation*} 
\bar{d} = \left( \prod^r_{i=1} d_i \right)^{1/r}
\end{equation*}
as the geometric mean of the singular values $d_1,..., d_r$.

This model is a closed-form expression model which allows for interpretation and straightforward calculation under any model. This is useful when calculating Bayes factors in terms of speed and efficiency since we do not have to use any MCMC estimation techniques. This gBF can be applied to all models, more specifically when the number of predictors $p$ exceeds the number of observations $n$. In our multivariate case, the design matrix \eqref{multixmatrix} will have $p > n$ dimensions.

\begin{theorem} \label{thm:surethm1}
(SURE for linear models). Let $Y\sim N(X\beta,\sigma^2 I_n)$, where the dimensions of $X$ are $n \times p$, and let $\hat{\beta} = \hat{\beta}(Y)$ be a weakly differentiable function of the least squares estimator $\hat{\beta}_{\text{OLS}}$ such that $\hat{Y} = X\hat{\beta}$ can be written in the form $\hat{Y}=a+SY$ for some vector \textbf{a} and matrix \textbf{S}. Let $\hat{\sigma}^2=||Y-X\hat{\beta}_{\text{OLS}}||^2/(n-p_*)$. Then,\begin{equation}
    \delta_0(Y) = ||Y-X\hat{\beta}||^2+(2\text{Tr}(S)-n)\hat{\sigma}^2
\end{equation}
is an unbiased estimator of $||Y-X\hat{\beta}||^2$.
\end{theorem}
\begin{theorem} \label{thm:surethm2}
(SURE minimization with respect to $g$). The value of $g$ that minimizes SURE in \eqref{resid} is \begin{equation}
    \label{gsure}
    g_* = \frac{||\hat{Y}_{\text{OLS}}-Y_0||^2}{p_* \hat{\sigma}^2}-1.
\end{equation}
\end{theorem}

%% file: 00-all.bbl
\begin{thebibliography}{34}
\providecommand{\natexlab}[1]{#1}
\providecommand{\url}[1]{\texttt{#1}}
\expandafter\ifx\csname urlstyle\endcsname\relax
  \providecommand{\doi}[1]{doi: #1}\else
  \providecommand{\doi}{doi: \begingroup \urlstyle{rm}\Url}\fi

\bibitem[Adekambi et~al.(2015)Adekambi, Ibegbu, Cagle, Kalokhe, Wang, Hu, Day, Ray, Rengarajan, et~al.]{adekambi2015biomarkers}
T.~Adekambi, C.~C. Ibegbu, S.~Cagle, A.~S. Kalokhe, Y.~F. Wang, Y.~Hu, C.~L. Day, S.~M. Ray, J.~Rengarajan, et~al.
\newblock Biomarkers on patient t cells diagnose active tuberculosis and monitor treatment response.
\newblock \emph{The Journal of Clinical Investigation}, 125\penalty0 (5):\penalty0 1827--1838, 2015.

\bibitem[Bartlett(1957)]{bartlett1957paradox}
M.~S. Bartlett.
\newblock A comment on {D.V. L}indley's statistical paradox.
\newblock \emph{Biometrika}, 44\penalty0 (3/4):\penalty0 533--534, 1957.

\bibitem[Bauer et~al.(2015)Bauer, Ahmed, Benedetti, Greenaway, Lalli, Leavens, Menzies, Vadeboncoeur, Vissandj{\'e}e, Wynne, et~al.]{bauer2015health}
M.~Bauer, S.~Ahmed, A.~Benedetti, C.~Greenaway, M.~Lalli, A.~Leavens, D.~Menzies, C.~Vadeboncoeur, B.~Vissandj{\'e}e, A.~Wynne, et~al.
\newblock Health-related quality of life and tuberculosis: a longitudinal cohort study.
\newblock \emph{Health and Quality of Life Outcomes}, 13:\penalty0 1--13, 2015.

\bibitem[Bayarri et~al.(2012)Bayarri, Berger, Forte, and Garcia-Donato]{bayarri2012criteria}
M.~J. Bayarri, J.~O. Berger, A.~Forte, and G.~Garcia-Donato.
\newblock Criteria for bayesian model choice with application to variable selection.
\newblock \emph{The Annals of Statistics}, 40\penalty0 (3):\penalty0 1550--1577, 2012.

\bibitem[Broll et~al.()Broll, Basu, Lee, and Wells]{PROLONG}
S.~Broll, S.~Basu, M.~H. Lee, and M.~T. Wells.
\newblock P{ROLONG}: Phenotype regression on longitudinal omics data with network and group lasso constraints.
\newblock URL \url{https://sumbose.stat.cornell.edu/wp}.

\bibitem[Casella(2001)]{10.1093/biostatistics/2.4.485}
G.~Casella.
\newblock {Empirical Bayes Gibbs sampling}.
\newblock \emph{Biostatistics}, 2\penalty0 (4):\penalty0 485--500, 12 2001.
\newblock ISSN 1465-4644.
\newblock \doi{10.1093/biostatistics/2.4.485}.
\newblock URL \url{https://doi.org/10.1093/biostatistics/2.4.485}.

\bibitem[Cliff et~al.(2013)Cliff, Lee, Constantinou, Cho, Clark, Ronacher, King, Lukey, Duncan, Van~Helden, Walzl, and Dockrell]{cliff_distinct_2013}
J.~M. Cliff, J.-S. Lee, N.~Constantinou, J.-E. Cho, T.~G. Clark, K.~Ronacher, E.~C. King, P.~T. Lukey, K.~Duncan, P.~D. Van~Helden, G.~Walzl, and H.~M. Dockrell.
\newblock Distinct {Phases} of {Blood} {Gene} {Expression} {Pattern} {Through} {Tuberculosis} {Treatment} {Reflect} {Modulation} of the {Humoral} {Immune} {Response}.
\newblock \emph{The Journal of Infectious Diseases}, 207\penalty0 (1):\penalty0 18--29, Jan. 2013.
\newblock ISSN 0022-1899.
\newblock \doi{10.1093/infdis/jis499}.
\newblock URL \url{https://doi.org/10.1093/infdis/jis499}.

\bibitem[Cornell et~al.(2014)Cornell, Mulrow, Localio, Stack, Meibohm, Guallar, and Goodman]{cornell2014random}
J.~E. Cornell, C.~D. Mulrow, R.~Localio, C.~B. Stack, A.~R. Meibohm, E.~Guallar, and S.~N. Goodman.
\newblock Random-effects meta-analysis of inconsistent effects: a time for change.
\newblock \emph{Annals of Internal Medicine}, 160\penalty0 (4):\penalty0 267--270, 2014.

\bibitem[Dutta et~al.(2020)Dutta, Tornheim, Fukutani, Paradkar, Tiburcio, Kinikar, Valvi, Kulkarni, Pradhan, Shivakumar, Kagal, Gupte, Gupte, Mave, Gupta, Andrade, and Karakousis]{dutta_integration_2020}
N.~K. Dutta, J.~A. Tornheim, K.~F. Fukutani, M.~Paradkar, R.~T. Tiburcio, A.~Kinikar, C.~Valvi, V.~Kulkarni, N.~Pradhan, S.~V. B.~Y. Shivakumar, A.~Kagal, A.~Gupte, N.~Gupte, V.~Mave, A.~Gupta, B.~B. Andrade, and P.~C. Karakousis.
\newblock Integration of metabolomics and transcriptomics reveals novel biomarkers in the blood for tuberculosis diagnosis in children.
\newblock \emph{Scientific Reports}, 10\penalty0 (1):\penalty0 19527, Nov. 2020.
\newblock ISSN 2045-2322.
\newblock \doi{10.1038/s41598-020-75513-8}.
\newblock URL \url{https://www.nature.com/articles/s41598-020-75513-8}.
\newblock Number: 1 Publisher: Nature Publishing Group.

\bibitem[Fourdrinier and Wells(2012)]{fourdrinier2012improved}
D.~Fourdrinier and M.~T. Wells.
\newblock On improved loss estimation for shrinkage estimators.
\newblock \emph{Statistical Science}, 27\penalty0 (1):\penalty0 61 -- 81, 2012.

\bibitem[Fourdrinier et~al.(2018)Fourdrinier, Strawderman, and Wells]{fourdrinier2018shrinkage}
D.~Fourdrinier, W.~E. Strawderman, and M.~T. Wells.
\newblock \emph{Shrinkage estimation}.
\newblock Springer, 2018.

\bibitem[Gomez et~al.(2016)Gomez, Dowdy, Bastos, Zwerling, Sweeney, Foster, Trajman, Islam, Kapiga, Sinanovic, et~al.]{gomez2016cost}
G.~Gomez, D.~W. Dowdy, M.~Bastos, A.~Zwerling, S.~Sweeney, N.~Foster, A.~Trajman, M.~Islam, S.~Kapiga, E.~Sinanovic, et~al.
\newblock Cost and cost-effectiveness of tuberculosis treatment shortening: a model-based analysis.
\newblock \emph{BMC Infectious Diseases}, 16:\penalty0 1--13, 2016.

\bibitem[Ha et~al.(2023)Ha, Chung, Bogardus, Jagannathan, Bergman, and Sherman]{ha2023one}
J.~Ha, S.~T. Chung, C.~Bogardus, R.~Jagannathan, M.~Bergman, and A.~S. Sherman.
\newblock One-hour glucose is an earlier marker of dysglycemia than two-hour glucose.
\newblock \emph{Diabetes Research and Clinical Practice}, 203:\penalty0 110839, 2023.

\bibitem[Isa et~al.(2018)Isa, Collins, Lee, Decome, Dorvil, Joseph, Smith, Salerno, Wells, Fischer, Bean, Pape, Johnson, Fitzgerald, and Rhee]{ISA2018157}
F.~Isa, S.~Collins, M.~H. Lee, D.~Decome, N.~Dorvil, P.~Joseph, L.~Smith, S.~Salerno, M.~T. Wells, S.~Fischer, J.~M. Bean, J.~W. Pape, W.~D. Johnson, D.~W. Fitzgerald, and K.~Y. Rhee.
\newblock Mass spectrometric identification of urinary biomarkers of pulmonary tuberculosis.
\newblock \emph{EBioMedicine}, 31:\penalty0 157--165, 2018.
\newblock ISSN 2352-3964.
\newblock \doi{https://doi.org/10.1016/j.ebiom.2018.04.014}.
\newblock URL \url{https://www.sciencedirect.com/science/article/pii/S2352396418301427}.

\bibitem[Johnstone and Silverman(2004)]{silverman2004}
I.~M. Johnstone and B.~W. Silverman.
\newblock {Needles and straw in haystacks: Empirical Bayes estimates of possibly sparse sequences}.
\newblock \emph{The Annals of Statistics}, 32\penalty0 (4):\penalty0 1594 -- 1649, 2004.
\newblock \doi{10.1214/009053604000000030}.
\newblock URL \url{https://doi.org/10.1214/009053604000000030}.

\bibitem[Kass and Raftery(1995)]{kass1995bayes}
R.~E. Kass and A.~E. Raftery.
\newblock Bayes factors.
\newblock \emph{Journal of the American Statistical Association}, 90\penalty0 (430):\penalty0 773--795, 1995.
\newblock \doi{10.1080/01621459.1995.10476572}.
\newblock URL \url{/brokenurl# http://www.tandfonline.com/doi/abs/10.1080/01621459.1995.10476572}.

\bibitem[Li et~al.(2021)Li, Ning, Li, Luo, Qin, Yu, Wang, Yang, Nan, He, Yang, Gong, Li, Liu, Sun, Li, Jia, Gao, Zhang, Huang, Hou, Xue, Li, Wei, Zhang, Li, and Wang]{li_longitudinal_2021}
T.~Li, N.~Ning, B.~Li, D.~Luo, E.~Qin, W.~Yu, J.~Wang, G.~Yang, N.~Nan, Z.~He, N.~Yang, S.~Gong, J.~Li, A.~Liu, Y.~Sun, Z.~Li, T.~Jia, J.~Gao, W.~Zhang, Y.~Huang, J.~Hou, Y.~Xue, D.~Li, Z.~Wei, L.~Zhang, B.~Li, and H.~Wang.
\newblock Longitudinal {Metabolomics} {Reveals} {Ornithine} {Cycle} {Dysregulation} {Correlates} {With} {Inflammation} and {Coagulation} in {COVID}-19 {Severe} {Patients}.
\newblock \emph{Frontiers in Microbiology}, 12, 2021.
\newblock ISSN 1664-302X.
\newblock URL \url{https://www.frontiersin.org/articles/10.3389/fmicb.2021.723818}.

\bibitem[Liang et~al.(2008)Liang, Paulo, Molina, Clyde, and Berger]{liang2008mixtures}
F.~Liang, R.~Paulo, G.~Molina, M.~A. Clyde, and J.~O. Berger.
\newblock Mixtures of g priors for {B}ayesian variable selection.
\newblock \emph{Journal of the American Statistical Association}, 103\penalty0 (481):\penalty0 410--423, 2008.

\bibitem[Lindley(1957)]{lindley1957paradox}
D.~V. Lindley.
\newblock A statistical paradox.
\newblock \emph{Biometrika}, 44\penalty0 (1-2):\penalty0 187--192, 1957.

\bibitem[Maruyama and George(2011)]{maruyama2011fully}
Y.~Maruyama and E.~I. George.
\newblock Fully {B}ayes factors with a generalized g-prior.
\newblock 2011.

\bibitem[Organization et~al.(2023)]{who2023}
W.~H. Organization et~al.
\newblock World {H}ealth {O}rganization {G}lobal {T}uberculosis {R}eport 2023.
\newblock \emph{URL: https://www.who.int/teams/global-tuberculosis-programme/tb-reports/global-tuberculosis-report-2023}, 2023.

\bibitem[Park and Casella(2008)]{doi:10.1198/016214508000000337}
T.~Park and G.~Casella.
\newblock The {B}ayesian lasso.
\newblock \emph{Journal of the American Statistical Association}, 103\penalty0 (482):\penalty0 681--686, 2008.
\newblock \doi{10.1198/016214508000000337}.
\newblock URL \url{https://doi.org/10.1198/016214508000000337}.

\bibitem[Porwal and Raftery(2022)]{doi:10.1073/pnas.2120737119}
A.~Porwal and A.~E. Raftery.
\newblock Comparing methods for statistical inference with model uncertainty.
\newblock \emph{Proceedings of the National Academy of Sciences}, 119\penalty0 (16):\penalty0 e2120737119, 2022.
\newblock \doi{10.1073/pnas.2120737119}.
\newblock URL \url{https://www.pnas.org/doi/abs/10.1073/pnas.2120737119}.

\bibitem[Ramel et~al.(2013)Ramel, Long, Gray, Durrwachter-Erno, Demerath, and Rao]{ramel2013neonatal}
S.~Ramel, J.~Long, H.~Gray, K.~Durrwachter-Erno, E.~Demerath, and R.~Rao.
\newblock Neonatal hyperglycemia and diminished long-term growth in very low birth weight preterm infants.
\newblock \emph{Journal of Perinatology}, 33\penalty0 (11):\penalty0 882--886, 2013.

\bibitem[Taheri et~al.(2020)Taheri, Zaghloul, Chagoury, Elhadad, Ahmed, El~Khatib, Amona, El~Nahas, Suleiman, Alnaama, Al-Hamaq, Charlson, Wells, Al-Abdulla, and Abou-Samra]{taheri2020diadem}
S.~Taheri, H.~Zaghloul, O.~Chagoury, S.~Elhadad, S.~H. Ahmed, N.~El~Khatib, R.~A. Amona, K.~El~Nahas, N.~Suleiman, A.~Alnaama, A.~Al-Hamaq, M.~Charlson, M.~T. Wells, S.~Al-Abdulla, and A.~B. Abou-Samra.
\newblock Effect of intensive lifestyle intervention on bodyweight and glycaemia in early type 2 diabetes (diadem-i): an open-label, parallel-group, randomised controlled trial.
\newblock \emph{The Lancet Diabetes \& Endocrinology}, 8\penalty0 (6):\penalty0 477--489, 2020.
\newblock \doi{10.1016/S2213-8587(20)30117-0}.

\bibitem[Venkatraman et~al.(2023)Venkatraman, Basu, Clark, Delbare, Lee, and Wells]{zellner}
S.~Venkatraman, S.~Basu, A.~G. Clark, S.~Delbare, M.~H. Lee, and M.~T. Wells.
\newblock An empirical {B}ayes approach to estimating dynamic models of co-regulated gene expression.
\newblock \emph{Data Science in Science}, 2\penalty0 (1):\penalty0 2219707, 2023.
\newblock \doi{10.1080/26941899.2023.2219707}.

\bibitem[Walsh et~al.(2020)Walsh, McAulay, Lee, Vilbrun, Mathurin, Jean~Francois, Zimmerman, Kaya, Zhang, Saito, Ocheretina, Savic, Dartois, Johnson, Pape, Nathan, and Fitzgerald]{walsh_early_2020}
K.~F. Walsh, K.~McAulay, M.~H. Lee, S.~C. Vilbrun, L.~Mathurin, D.~Jean~Francois, M.~Zimmerman, F.~Kaya, N.~Zhang, K.~Saito, O.~Ocheretina, R.~Savic, V.~Dartois, W.~D. Johnson, J.~W. Pape, C.~Nathan, and D.~W. Fitzgerald.
\newblock Early {Bactericidal} {Activity} {Trial} of {Nitazoxanide} for {Pulmonary} {Tuberculosis}.
\newblock \emph{Antimicrobial Agents and Chemotherapy}, 64\penalty0 (5):\penalty0 e01956--19, Apr. 2020.
\newblock ISSN 1098-6596.
\newblock \doi{10.1128/AAC.01956-19}.

\bibitem[Wipperman et~al.(2021)Wipperman, Bhattarai, Vorkas, Maringati, Taur, Mathurin, McAulay, Vilbrun, Francois, Bean, Walsh, Nathan, Fitzgerald, Glickman, and Bucci]{wipperman_gastrointestinal_2021}
M.~F. Wipperman, S.~K. Bhattarai, C.~K. Vorkas, V.~S. Maringati, Y.~Taur, L.~Mathurin, K.~McAulay, S.~C. Vilbrun, D.~Francois, J.~Bean, K.~F. Walsh, C.~Nathan, D.~W. Fitzgerald, M.~S. Glickman, and V.~Bucci.
\newblock Gastrointestinal microbiota composition predicts peripheral inflammatory state during treatment of human tuberculosis.
\newblock \emph{Nature Communications}, 12\penalty0 (1):\penalty0 1141, Feb. 2021.
\newblock ISSN 2041-1723.
\newblock \doi{10.1038/s41467-021-21475-y}.
\newblock URL \url{https://www.nature.com/articles/s41467-021-21475-y}.
\newblock Number: 1 Publisher: Nature Publishing Group.

\bibitem[Wooldridge(2010)]{wooldridge2010econometric}
J.~M. Wooldridge.
\newblock \emph{Econometric Analysis of Cross-Section and Panel Data}.
\newblock MIT press, 2010.

\bibitem[Xia et~al.(2020)Xia, Lee, Walsh, McAulay, Bean, Fitzgerald, Dupnik, Johnson, Pape, Rhee, and Isa]{10.1172/jci.insight.136301}
Q.~Xia, M.~H. Lee, K.~F. Walsh, K.~McAulay, J.~M. Bean, D.~W. Fitzgerald, K.~M. Dupnik, W.~D. Johnson, J.~W. Pape, K.~Y. Rhee, and F.~Isa.
\newblock Urinary biomarkers of mycobacterial load and treatment response in pulmonary tuberculosis.
\newblock \emph{JCI Insight}, 5\penalty0 (18), 9 2020.
\newblock \doi{10.1172/jci.insight.136301}.
\newblock URL \url{https://insight.jci.org/articles/view/136301}.

\bibitem[Xu and Ghosh(2015)]{xu2015bayesian}
X.~Xu and M.~Ghosh.
\newblock Bayesian variable selection and estimation for group lasso.
\newblock 2015.

\bibitem[Yuan and Lin(2006)]{yuan2006model}
M.~Yuan and Y.~Lin.
\newblock Model selection and estimation in regression with grouped variables.
\newblock \emph{Journal of the Royal Statistical Society Series B: Statistical Methodology}, 68\penalty0 (1):\penalty0 49--67, 2006.

\bibitem[Zellner(1986)]{Zellner1986}
A.~Zellner.
\newblock \emph{On assessing prior distributions and Bayesian regression analysis with g-prior distributions}.
\newblock Elsevier, New York, NY, 1986.

\bibitem[Zhang et~al.(2014)Zhang, Baladandayuthapani, Mallick, Manyam, Thompson, Bondy, and Do]{10.1111/rssc.12053}
L.~Zhang, V.~Baladandayuthapani, B.~K. Mallick, G.~C. Manyam, P.~A. Thompson, M.~L. Bondy, and K.-A. Do.
\newblock {Bayesian Hierarchical Structured Variable Selection Methods with Application to Molecular Inversion Probe Studies in Breast Cancer}.
\newblock \emph{Journal of the Royal Statistical Society Series C: Applied Statistics}, 63\penalty0 (4):\penalty0 595--620, 03 2014.
\newblock ISSN 0035-9254.
\newblock \doi{10.1111/rssc.12053}.
\newblock URL \url{https://doi.org/10.1111/rssc.12053}.

\end{thebibliography}
